\documentclass[eqsecnum,aps,preprint,epsf ]{revtex4}
\usepackage{graphicx}
\usepackage{wrapfig}
\font\tenbifull=cmmib10 scaled 1200 % bold math italic
\font\tenbimed=cmmib9
\font\tenbismall=cmmib7
\def\bmit{\fam9 }
\mathchardef\bbkappa="7114
\mathchardef\bbrho="711A
\mathchardef\bbsigma="711B
\mathchardef\bbtau="711C
\mathchardef\bbvarrho="7125
\mathchardef\bbvarsigma="7126
\mathchardef\bbPhi="7008
\mathchardef\bbxi="7118

\def\boldsigma{{\bmit\bbsigma}}
\def\boldtau{{\bmit\bbtau}}

\def\boldPhi{{\bmit\bbPhi}}
\def\boldxi{{\bmit\bbxi}}
\newcommand {\CM } {{\cal M}}
\newcommand {\la} {\left\langle}
\newcommand {\ra} {\right\rangle}
\newcommand{\be}{\begin{eqnarray}&&}
\newcommand{\ee}{\end{eqnarray}}

\def\bsp{ \left | {\bf p}_{\rm sp} \right |}
\def\bpD{ \left | {\bf P}_{D} \right |}
\begin{document}
\title{Di-Electron Bremsstrahlung in Intermediate-Energy \mbox{\boldmath $pn$}  and
\mbox{\boldmath $Dp$}  Collisions}

\author{L.P. Kaptari}
\altaffiliation{On leave of absence from
 Bogoliubov Lab. Theor. Phys. 141980, JINR,  Dubna, Russia}
\author{ B. K\"ampfer}
\affiliation{Forschungszentrum Rossendorf, PF 510119, 01314 Dresden, Germany}

\date{\today}

\begin{abstract}
Invariant mass spectra of di-electrons stemming from bremsstrahlung processes
are calculated in a covariant diagrammatical  approach
for the exclusive reaction $D p \to p_{\rm sp} \, n p \: e^+ e^-$
with detection of a forward spectator proton, $p_{\rm sp}$.
We employ an effective nucleon-meson theory for parameterizing the
sub-reaction $n p \to n p \: e^+ e^-$
and, within the Bethe-Salpeter formalism, derive a factorization of the cross section
in the form
${d\sigma_{D p \to p_{\rm sp} \, n p \: e^+ e^-}}/{dM}=
{d\sigma_{n p \to  \, n p \: e^+ e^-}}/{dM}\times$~kinematical
factor related solely to the deuteron ($M$ is the $e^+ e^-$ invariant mass).
The effective nucleon-meson interactions, including the exchange
mesons $\pi$, $\sigma$, $\omega$ and $\rho$ as well as excitation and radiative
decay of $\Delta(1232)$,  have been adjusted to the process
$pp \to pp \: e^+ e^-$ at energies below the vector meson production threshold.
At higher energies, contributions from $\omega$ and $\rho$ meson excitations
are analyzed in both, $NN$ and $Dp$ collisions. A relation
to two-step models is  discussed.
Subthreshold di-electron production in $Dp$ collisions
at low spectator momenta is investigated as well. Calculations have been
performed for kinematical conditions envisaged for forthcoming experiments at
HADES.
\end{abstract}
\maketitle

\section{Introduction}

Di-electron production in scattering processes of hadrons
at low energies can be described essentially
as  bremsstrahlung from  incoming and outgoing  charged particles.
Several formulae for bremsstrahlung, obtained within different
approximations, have been proposed (for a survey of theoretical approaches to
bremsstrahlung reactions see e.g.\ \cite{tormoz}).
With the focus on intermediate energies,
a covariant approach based on an effective meson-nucleon theory to calculate
the bremsstrahlung of di-electrons from nucleon-nucleon scattering
has recently been presented in \cite{mosel_calc},
continuing and extending the series of previous investigations
\cite{Kapusta}. In this model
the effective parameters have been adjusted to describe
elastic nucleon-nucleon ($NN$) and inelastic $NN \to NN \pi$
processes at intermediate energies; besides, the role
of  excitations of intermediate resonances
has been studied within
this approach and it is found that at intermediate energies the main contribution
comes from $\Delta$ resonances
(see also Ref.~\cite{ernst}), whereas excitations of higher mass
resonances can be neglected.
The role of higher mass and spin nucleon resonances
at energies near the vector meson ($\rho$, $\omega$ and $\phi$)
production thresholds have been investigated in some
detail for proton-proton collisions
in several papers (see, e.g., Refs.~\cite{nakayama,fuchs} and references therein quoted)
with the conclusion that at threshold-near energies the inclusion
of heavier resonances also leads to good description of data.
However, as demonstrated in Refs.~\cite{nakayama,ourOmega}
calculations with a reasonable readjustment of the effective parameters
can equally well describe the data without higher mass and spin resonances.
In contrast, for di-electron production in photon and pion
induced reactions excitations of low-lying as well as
heavier resonances can play a role \cite{lutzNew}.

In the present paper we extend the covariant model \cite{mosel_calc,ourOmega},
which is based on an effective meson nucleon theory with inclusion of $\Delta$ isobar
contributions and  vector meson dominance,
to the exclusive process $D p \to p_{\rm sp}\ n p\   e^+ e^-$
with incoming deuteron $D$ where the di-electron $e^+ e^-$ is
detected in coincidence with the fast spectator proton $p_{\rm sp}$.
We calculate the cross section of di-electrons produced primarily in
bremsstrahlung processes which, to some extent, can be considered as
background contribution to other, more complicate processes. In order to preserve
the covariance of the approach, the deuteron ground state and the corresponding
matrix elements are treated within the Bethe-Salpeter (BS) formalism by making use
of a realistic solution of the BS equation, obtained within the same
effective meson nucleon theory \cite{solution}.

Corresponding experiments are planned by the HADES collaboration at the
heavy ion synchrotron SIS/GSI Darmstadt \cite{HADES}.
The outgoing neutron $n$ and proton $p$ may be reconstructed by the
missing mass technique. The very motivation of such experiments is to pin down the bremsstrahlung
component for $e^+ e^-$ production in the tagged sub-reaction
$n p \to n p\: e^+ e^-$ \cite{holzman}.
Detailed knowledge of this reaction, together with the directly measurable
reaction $p p \to p p\: e^+ e^-$, is a necessary prerequisite for understanding
di-electron emission in heavy-ion collisions.
In heavy-ion collisions the di-electron  rate
is determined by the retarded photon self-energy in medium, which in turn is
related to in-medium propagators of various vector mesons.
In such a way the in-medium modifications of vector mesons become directly accessible.

A broader scope is to achieve a refined understanding of the nucleon-nucleon force
at intermediate energies. Similar to meson production or real bremsstrahlung,
the virtual bremsstrahlung processes probe some off-shell part of the amplitude
providing more  profound  insights into the electromagnetic structure of hadrons,
e.g., the electromagnetic form factors in the time-like
region, not accessible  in on-mass shell reactions.
Another important issue of  di-electron emission in $NN$ collisions is to
supply additional information on vector meson production, in particular
$\omega$ and $\phi$ mesons \cite{ourOmega,ourphi}, which is interesting in respect  to
the Okubo-Zweig-Iizuka rule \cite{ozi} and hidden strangeness in the nucleon.

Our paper is organized as follows. In section II we consider di-electron production
in elementary $NN$ collisions by parameterizing the amplitudes by corresponding Feynman
diagrams. Parameters are adjusted to $pp$ collisions.
In section III we extend the approach to
deuteron-proton collisions by employing the Bethe-Salpeter formalism
with a realistic solution obtained with  one-boson-exchange kernel.
Within the spectator mechanism picture we derive a factorization formula
relating the reactions  $D p \to p_{\rm sp} p n \: e^+ e^-$ and $n p \to n p \: e^+ e^-$.
The summary and discussion can be found in section IV.
Compilations of useful formulae are summarized in Appendices \ref{isospin} and \ref{ts}
(where a link to two step models is outlined),
while the Appendix \ref{a2} describes some
details needed for deriving the factorization formula.

\section{Di-electrons from \mbox{\boldmath $NN $} collisions}
\label{section1}

\subsection{Kinematics and Notation}  \label{subsection1}

We consider the exclusive $e^+e^-$ production in $NN$ reactions of the type
\be
N_1 (P_1) + N_2 (P_2)  \to N_1' (P_1') + N_2' (P_2') + e^+(k_1) + e^-(k_2)
\label{reac1}
\ee
(For an extension towards including hadronic inelasticities
in semi-inclusive reactions cf.\ \cite{Haglin_Gale}.) 
The invariant eight-fold cross section is
\begin{equation}
d^8\sigma =
\frac{1}{2\sqrt{\lambda(s,m^2,m^2)}}
\frac14 \sum\limits_{\rm spins}\
\,|\ T(P_1',P_2',k_1,k_2,{\rm spins}) \ |^2 d^8\tau_f \ \frac{1}{n! },
\label{crossnn}
\end{equation}
where the kinematical  factor $\lambda$ is
$\lambda(x^2,y^2,z^2)=(x^2-(y+z)^2)(x^2-(y-z)^2)$;
the factor $1/n!$ accounts for $n$ identical
particles in the final state, $\vert T \vert^2$ denotes the invariant amplitude squared.
The invariant phase space volume $d\tau_f$ is defined as
\begin{widetext}
\begin{equation}
d^8\tau = (2\pi)^4\delta\left(P_1+P_2-P_1'-P_2'-k_1-k_2\right)
\prod
\frac{d^3P_i'}{2E_{{\bf p}_i'}(2\pi)^3}
\prod
\frac{d^3k_i}{2E_{{\bf k}_i}(2\pi)^3}.
\label{tau}
\end{equation}
\end{widetext}
The 4-momenta
of initial  ($P_1, P_2$) and  final  ($P_1', P_2'$) nucleons are
$P=(E_{\bf P},{\bf P})$ with $E_{\bf P}=\sqrt{m^2+{\bf P}^2}$, an  analogous notation
is used for the lepton momenta $k_{1,2}$;
$m$ denotes the nucleon mass, while the electron mass
can be neglected for the present kinematics.
The invariant mass of two particles is hereafter denoted as $s$ with $s=(P_1+P_2)^2$;
along with this notation  for the invariant mass
of the virtual photon throughout the paper we also use
the more familiar notation $q^2$  with $q^2 \equiv s_\gamma$.
As  seen from (\ref{tau}), the cross section
Eq.~(\ref{crossnn}) is determined by eight independent kinematical variables, the
actual choice of which depends upon the specific goals of the considered problem.
In the present paper we are mainly  interested in studying the
invariant mass distribution of the produced electrons and positrons. For this sake
it is convenient to choose the kinematics with two invariants,
$s_\gamma = (k_1 + k_2)^2$  and $s_{12}=(P_1'+P_2')^2$, and three solid
angles, $d\Omega_\gamma^* $,  $d\Omega_{12}^*$ and $d\Omega_{\pm}^*$.
This corresponds to creation of two intermediate particles with invariant masses
$\sqrt{s_\gamma}$ and $\sqrt{s_{12}}$ with their subsequent decay
into two final nucleons
and two leptons, respectively, as depicted in Fig.~\ref{fig1}.
For each pair, the kinematical variables will be defined
in the corresponding two-particle center-of-mass (CM) system. This can be achieved, e.g.,
by inserting in Eq.~(\ref{tau}) the identities
\begin{eqnarray}
1&=&\int ds_\gamma \ d^4 P_{\gamma} \ \delta (P_{\gamma}^2-s_\gamma )\
\delta^{(4)} (P_{\gamma}-k_1-k_2),  \label{unity}\\
1&=&\int ds_{12} \ d^4 P_{12}\  \delta (P_{12}^2-s_{12} )
\delta^{(4)} (P_{12}-P_1'-P_2')
\label{odin1}
\end{eqnarray}
and rearranging terms in Eq.~(\ref{tau}) to separate the invariant phase space
volumes for the ''decays'' with $P_\gamma = k_1 + k_2$ and
$P_{12} = P_1' + P_2'$, see Fig.~\ref{fig1}.
With these conventions we arrive at
\begin{widetext}
\begin{eqnarray} % \raggedright
d^8\sigma &=&
\frac{1}{2\sqrt{\lambda(s,m^2,m^2)}}\frac{1}{(2\pi)^8}
 \frac14\sum\limits_{spins} \,
|T |^2 \,  \frac{1}{n! } d s_{12} ds_\gamma \\
&\times&
R_2(P_\gamma\to k_1+k_2)\ R_2(P_{12}\to P_1'+P_2')\ R_2(P_1+P_2\to P_\gamma+P_{12}),
\nonumber
\label{sechenie}
\end{eqnarray}
\end{widetext}
where the two-body invariant phase space volume $R_2$ is defined as
\begin{equation}
R_2(a+b \to c+d) = d^4 P_c\ d^4P_d\ \delta^{(4)}(P_a+P_b-P_c-P_d)\
\delta (P_c^2-m_c^2)\ \delta(P_d^2-m_d^2).
\label{spase}
\end{equation}

\subsection{Leptonic tensor}

In the lowest order of the electromagnetic coupling
(one-photon approximation) the di-electron
production process is considered as decay of a virtual photon
produced in strong and electromagnetic $NN$ interactions
from different elementary reactions, e.g.,
bremsstrahlung, Dalitz decay, vector meson decay etc.\ \cite{brat}.
For such a process the general expression for the
invariant amplitude squared reads
\be
|T|^2=W_{\mu\nu} \, \frac{e^4}{q^4}  l^{\mu\nu},
\label{ophe}
\ee
where the momentum $P_\gamma$ of the virtual photon is denoted as
$q \equiv P_\gamma=(k_1+k_2)$;  $e$ is the elementary charge.
The purely electromagnetic decay vertex of the virtual photon is
determined by the leptonic tensor
$ l^{\mu\nu} =\sum\limits_{spins} j^\mu j^\nu$
with the current
$j^\mu = \bar u(k_1,s_1)\ \gamma^\mu v (k_2,s_2)$,
where $\bar u$ and $v$ are the corresponding Dirac wave functions for the outgoing
electron and positron. For unpolarized di-electrons the leptonic tensor reads explicitly
\be
l_{\mu\nu} = 4 \left( k_{1\mu} k_{2\nu}+  k_{1\nu} k_{2\mu}
-g_{\mu\nu} (k_1\cdot  k_2)\right).
\ee
The electromagnetic hadronic current $J_\mu$ and  the hadronic tensor
$W_{\mu\nu}=\sum\limits_{spins} J_\mu J_\nu^+$, besides 
the electromagnetic interaction, also involve
the strong interaction between the interacting nucleons and, consequently,
are of a  more complicate nature than $j_\mu$ and $l_{\mu\nu}$.
In virtue of gauge invariance the electromagnetic tensors obey
$q_\mu l^{\mu\nu} = q_\nu l^{\mu\nu} = q^\nu W_{\mu\nu} = q^\nu W_{\mu\nu}=0$, and
one can omit in $ l^{\mu\nu}$ all terms proportional to $q_\mu$ and $q_\nu$ and write
\be
{l}_{\mu\nu} = -2(4k_{1\mu}  k_{1\nu}+s_\gamma g_{\mu\nu} ).
\label{leptensor}
\ee
It is evident that the leptonic tensor depends solely upon the kinematical
variables connected  with the virtual photon vertex (see Fig.~\ref{fig1}) and
is independent of the variables determining the nucleon-nucleon interaction.
This implies that, due to Lorentz invariance of both the amplitude Eq.~(\ref{ophe})
and the corresponding phase space volume $R_2(P_\gamma\to k_1+k_2)$,
one can carry out the integration over the leptonic
variables in any system of reference.
The integration is particularly simple in the CM of the leptonic pair, where
${\bf q} = 0$,
$R_2=\sqrt{\lambda(s_\gamma,\mu_e^2,\mu_e^2)}/8s_\gamma \ d\Omega_{\pm}$ and
all the time like components of $l_{\mu\nu}$ vanish.  One obtains
\begin{eqnarray}&&
\int d\Omega_{\pm}^*
W_{\mu\nu} \frac{e^4}{q^4} l^{\mu\nu} =
-\frac{16\pi e^4}{3}\frac{J_\mu J^{ +\mu}}{s_\gamma}.
\end{eqnarray}
The remaining integrals can be computed by evaluating each
differential volume $R_2$  also in
the corresponding CM system. All together we obtain
\begin{eqnarray}
\frac{d\sigma}{d s_\gamma}=
-\frac{\alpha_{em}^2}{12s s_\gamma (4\pi)^5}
\int d s_{12} d\Omega_\gamma^* d\Omega_{12}^*
\sqrt{\displaystyle\frac{\lambda(s,s_\gamma,s_{12})\lambda(s_{12},m^2,m^2)}
{ s_{12}^2\lambda(s,m^2,m^2)}} \sum\limits_{spins} J_\mu J^{  +\mu},
\label{diffcross}
\end{eqnarray}
where $d\Omega_\gamma^*$
and   $d\Omega_{12}^*$ are defined in the CM of initial and final nucleons, respectively;
$\alpha_{em}$ stands for the electromagnetic fine structure constant.

\subsection{Lagrangians and parameters}

The covariant hadronic current $J_\mu$ is evaluated
within a meson-nucleon theory based on effective interaction Lagrangians
which consist on two parts, describing the strong and electromagnetic
interaction. In our approach, the strong interaction
among nucleons is mediated by four exchange mesons:
scalar ($\sigma$), pseudoscalar-isovector  ($\pi$),
and neutral vector ($\omega$) and vector-isovector  ($\rho$)  mesons
\cite{mosel_calc,ourOmega,ourPhi,scyam_lagr,bonncd,chung}.
We adopt the nucleon-nucleon-meson (NNM)  interaction terms
\begin{eqnarray}
{\cal L}_{NN\sigma }&=& g_\sigma \bar N  N \it\Phi_\sigma , \\
{\cal L}_{ NN\pi}&=&
-\frac{f_{ NN\pi}}{m_\pi}\bar N\gamma_5\gamma^\mu \partial_\mu
({\boldtau \boldPhi_\pi})N , \label{pseudo_vector}\\
{\cal L}_{ NN\rho}&=&
-g_{ NN \rho}\left(\bar N \gamma_\mu{\boldtau}N{\boldPhi_ \rho}^\mu-\frac{\kappa_\rho}{2m}
\bar N\sigma_{\mu\nu}{\boldtau}N\partial^\nu{\boldPhi_\rho}^\mu\right) , \\
{\cal L}_{ NN\omega}&=&
-g_{  NN \omega}\left(
\bar N \gamma_\mu N {\it\Phi}_{\omega}^\mu-
\frac{\kappa_{\omega}}{2m}
\bar N \sigma_{\mu\nu}  N \partial^\nu \it\Phi_{\omega}^\mu\right),
\label{mnn}
\end{eqnarray}
where $N$ and $\it\Phi_M$ denote the nucleon and meson fields, respectively,
and bold face letters stand for isovectors.
All couplings with off-mass shell
particles are dressed by monopole form factors
$F_M=\left(\Lambda^2_M-\mu_M^2\right)/\left(\Lambda^2_M-k^2_M\right)$,
where $k^2_M$ is the 4-momentum of a virtual particle
with mass $\mu_M$.
The effective parameters are adjusted to  experimental data
on $N N$ scattering. At low energies (below the pion threshold)
these parameters are rather well known 
and can be taken from iterated ${\cal T}$ matrix fits of experimentally
known elastic phase shifts \cite{bonncd}. At intermediate energies,
say in the interval 1 - 3 $GeV$,
it turns out that the pure tree level description basing on the above interactions
is not able to reproduce equally well the  energy dependence of the
data \cite{mosel_calc,ikh_model}.
Therefore, following \cite{mosel_calc,ikh_model}, we take into account
an  energy dependence of the effective couplings
\be
g_{NNM} \to g_{NNM}(s)=g_0 {\rm e}^{-l \sqrt{s}}.
\label{obeMosel}
\ee
In what follows we employ the
parameters $l, g_0$ and $\Lambda_M$  from \cite{mosel_calc} which assure
a good tree level description of the elastic $NN \to NN$, $NN \to N \Delta$
and inelastic $NN \to NN\pi$ reactions at intermediate energies.
Note that problems with double counting of the $\Delta$ degrees of freedom
are avoided in such an prescription.

\subsection{Nucleon form factors and gauge invariance}\label{subsecGauge}

The form of the cross section Eq.~(\ref{diffcross})
bases essentially on the gauge invariance of hadronic and leptonic
tensors. This implies that in elaborating  models
for the reaction (\ref{reac1}) with effective Lagrangians
particular attention must be devoted to
the gauge invariance of the computed currents.
In our approach, i.e., in
the one-boson exchange approximation (OBE) for the strong $NN$ interaction
and one-photon exchange for the electromagnetic
production of $e^+e^-$, the current $J_\mu$ is determined by diagrams of two types:
(i) the ones which describe the creation of a virtual photon
with $q^2>0$ as pure nucleon bremsstrahlung as depicted in Fig.~\ref{fig2} and 
\ref{fig3}b, c, and
(ii) in case of exchange of charged mesons, emission of a virtual $\gamma^*$
from internal meson lines, see Fig.~\ref{fig3}a.
For these diagrams the gauge invariance is tightly connected with the
two-body Ward-Takahashi (WT) identity (see \cite{origwt,wt1,wt2,wtTjon}
and further references therein quoted)
\be
q_\mu \Gamma^\mu (p',p) = \frac{e(1+\tau_3)}{2}\left(
S^{-1}(p') - S^{-1}(p)\right),
\label{wt}
\ee
where $\Gamma^\mu$ is the electromagnetic vertex and
$S(p)$ is the (full) propagator of the particle. It is straightforward to
show that, if (\ref{wt}) is to be fulfilled, then
pairwise two diagrams with exchange of neutral
mesons and  pre-emission and post-emission of $\gamma^*$ (cf.\  Fig.~\ref{fig2}b))
cancel each other, hence ensuring $q^\mu J_\mu=0$, i.e.,
current conservation (see \cite{vmd_ff}).
This is also true even after dressing the vertices with phenomenological
form factors. However, in case of charged meson exchange
the WT identity is not any more automatically fulfilled. This is
because the nucleon momenta are interchanged
and, consequently,  the "right" and "left" internal nucleon propagators are defined
for different momenta of the exchanged meson. For instance, the contribution
to  $q^\mu J_\mu$ from the bremsstrahlung diagrams
Fig.~\ref{fig2}a) and \ref{fig2}d) reads
\begin{eqnarray}&&
%q_\mu\ J^{\mu} \sim \nonumber\\&&
\bar u(P_1')\Gamma_{NNM} S\left(P_1-q\right)
\left[ S^{-1}\left(P_1-q\right) - S^{-1}(P_1)\right] u(P_1)
\frac{i}{\left( k_1^2-\mu_M^2\right)} \bar u(P_2')\Gamma_{NNM} u(P_1) +
\nonumber\\&&
\bar u(P_1')\Gamma_{NNM} u(P_1) \frac{i}{\left( k_2^2-\mu_M^2\right)} \bar u(P_2')
 \left[ S^{-1}\left(P_2'\right) - S^{-1}(P_2'+q)\right] S\left(P_2'+q\right)
 \Gamma_{NNM} u(P_2)\nonumber\\&&
= \bar u(P_1')\Gamma_{NNM} u(P_1)\left[
\frac{i}{\left( k_1^2-\mu_M^2\right)}
-\frac{i}{\left( k_2^2-\mu_M^2\right)}\right] \bar u(P_2')\Gamma_{NNM} u(P_2),
%\neq 0,
\label{iso}
\end{eqnarray}
where $k_1=P_2'-P_2$ , $ k_2=P_1'-P_1$, $S^{-1}\left(P_1\right) u(P_1)=0$ and
$\bar u(P_2') S^{-1}\left(P_2'\right)=0$.
In fact, $q_\mu J^\mu \ne 0$ follows.
In order to restore the gauge invariance on this level
one must consider additional diagrams with emission of the
virtual photon by the charged meson exchange as depicted in Fig.~\ref{fig3}a. 
Then it is easy to show that the contribution from this diagram exactly compensates
the non-zero part (\ref{iso}),
and thus gauge invariance is restored.
This holds for bar $NNM$ vertices without cut-off form factors.
Inclusion of additional form factors again leads to non-conserved
currents. There are several prescriptions of how
to preserve gauge invariance within effective theories with
cut-off form factors \cite{wtTjon,mosel_calc,vmd_ff,gross}. The main idea of these
prescriptions is to include the cut-off form factors into WT identity
explicitly  and to consider the new relations as the WT identity for the full
propagators. For instance, Refs~\cite{vmd_ff,gross}  suggest to interpret
the cut-off form factors as an effective account of
the self-energy corrections and to present the full propagators, entering the WT
identity (\ref{wt}), as the bare ones multiplied from the both ends
of the propagator line by phenomenological cut-off functions.
Formally, all the Feynman rules to calculate  ladder diagrams as exhibited in
Fig.~\ref{fig2} remain unchanged, while in calculation of diagram types as depicted in
Fig.~\ref{fig3}a the meson-nucleon vertex must  be multiplied by the square of the
cut-off form factor $F_{MNN}^2(k)$; consequently a factor  $F_{MNN}^{-1}(k)$
must be included into the effective electromagnetic $MM\gamma$  vertex.
In the simplest case the bare mesonic vertex $\Gamma_\mu^M=\left( k_{1\mu}+k_{2\mu}\right)$
receives an additional factor \cite{schafer,mathiot}
\be
F_{add}=1-\frac{k_1^2-\mu_M^2}{\Lambda_m^2-k_2^2}
-\frac{k_2^2-\mu_M^2}{\Lambda_M^2-k_1^2}
\label{addmec}
\ee
becoming
\be
\Gamma_\mu^{\gamma M}=\left( k_{1\mu}+k_{2\mu}\right)
\frac{\left(\Lambda_M^2 - k_1^2\right)}{\left(\Lambda_M^2 - \mu_M^2\right)}
\frac{\left(\Lambda_M^2 - k_2^2\right)}{\left(\Lambda_M^2 - \mu_M^2\right)}
F_{add}.
\label{pionem}
\ee
It can be seen that the "renormalized"  vertex (\ref{pionem}) obeys the WT identity
for the full, renormalized mesonic propagators. In a more general case,
on can add to the vertex (\ref{pionem}) any divergenceless term, which obviously does not
change the WT identity.  Often it is convenient to
display  in the mesonic vertices some terms  which assure the WT identity
and the  divergenceless part explicitly, in which case the
corresponding vertex reads as (see also Ref.~\cite{wtTjon})
\begin{eqnarray} &&
\Gamma_\mu^{\gamma M} = \frac{q_\mu}{q^2}\left(\Delta^{-1}(k_1^2) - \Delta^{-1}(k_2^2)\right)
     +B(k_1,k_2)
     \left[(k_{1\mu}+k_{2\mu}) - q_\mu\frac{  q\cdot(k_1+k_2)  }{q^2}\right],
     \label{moeWT}
\end{eqnarray}
where $\Delta(k^2)$ denotes the scalar propagator and $B(k_1,k_2)$ 
is an arbitrary scalar function.
In accordance with \cite{gross}, the meson propagators are to be multiplied
by cut-off form factors
at both ends of their  lines in the diagram Fig.~\ref{fig3}a, resulting in
\be
\Delta (k^2) =\frac{F_{MNN}^2(k^2)}{k^2-\mu_M^2}.
\ee

Note that the above prescriptions for restoration of the gauge invariance
in $pn$ collisions are valid only if the effective meson-nucleon
interaction vertices do not depend on the momentum $k$ of the exchanged meson.
This is the case for pseudo-scalar $\pi NN $ coupling. Instead, if
the pseudo-vector $\pi NN $ coupling (\ref{pseudo_vector}) is chosen then
the Fourier transformed four divergence of the currents corresponding to
diagrams Fig.~\ref{fig2}  contains an additional
$ k$ dependence from the derivatives in the $\pi NN$ vertex
(cf.\ Eq.~(\ref{mnn})) yielding
\begin{eqnarray}
\!\!\!\!\!\!\!\!
q_\mu J^\mu \sim  \bar u(p_1') \gamma_5 \hat k_1 u(p_1)
\Delta(k_1^2)  \bar u(p_2') \gamma_5 \hat k_1 u(p_2)%\nonumber\\&&
- \bar u(p_1') \gamma_5 \hat k_2 u(p_1)
\Delta(k_2^2)  \bar u(p_2') \gamma_5 \hat k_2 u(p_2).
\label{WTNu}
\end{eqnarray}
The contribution  of the diagram  Fig.~\ref{fig3}a with
the mesonic vertex (\ref{pionem}) or (\ref{moeWT}) to $q^\mu J_\mu$ reads very similar to
(\ref{WTNu}) but in each term  both momenta, $\hat k_1$ and $\hat k_2$ enter, and
consequently, the gauge invariance can not be completely restored.
Within an effective meson nucleon theory
with  interaction Lagrangians depending on
derivatives, the gauge invariant coupling with photons
is introduced by replacing partial derivatives,
including the $NN M$ vertices, by a gauge covariant form (minimal coupling).
Such a procedure generates another kind of Feynman diagrams with
contact terms, i.e., vertices with four lines, known also as
Kroll-Rudermann \cite{KR} or seagull like diagrams, see Figs.~\ref{fig3}b and c.
We include therefore in our calculations these diagrams and the corresponding
interaction Lagrangian
\begin{equation}
{\cal L}_{ NN\pi\gamma} =
-\frac{\hat e f_{ NN\pi}}{m_\pi}\bar N\gamma_5\gamma^\mu A_\mu
({\boldtau \boldPhi_\pi})N
\end{equation}  
with 4-potential $A_\mu$ and charge operator $\hat e$ of the pion.
Gauge invariance is henceforth ensured. The contact term contributions
become particularly important near the kinematical limits. 

All electromagnetic $NN\gamma$ vertices corresponding to the interaction
Lagrangian
\begin{eqnarray}
{\cal L}^{em}_{NN\gamma}= -e\left( \bar N\gamma_\mu N \right) A^\mu+
e\kappa \bar N \left( \frac{\sigma_{\mu\nu}}{4m}{\cal F}^{\mu\nu}\right) N
\label{elm}
\end{eqnarray}
with the field strength tensor
${\cal F}^{\nu\mu}=\partial_\nu A^\mu - \partial_\mu A^\nu$,
and $\kappa$ as the anomalous magnetic moment of the nucleon
($\kappa=1.793$ for  protons and $\kappa=-1.913$ for neutrons),
should  also be dressed by  form factors.
This means that at least two form factors are needed 
to describe electron scattering from on-mass shell nucleons.

In the more general case of off-mass shell nucleons even eight terms, 
satisfying the necessary
symmetry and gauge invariance requirements, with eight scalar
form factors contribute to the $NN\gamma$ vertex. At $q^2 < 0 $ and low nucleon
virtuality ($P^2\sim m^2$) it is still possible to restrict this set to  two
effective form factors to describe electron scattering from off-mass shell
nucleons (e.g., in  $A(e,e'p)$ reactions) by modifying (kinematically)
the $NN\gamma$ vertex to satisfy  gauge invariance (see for details,
\cite{deForest} and further references therein quoted). Unfortunately, for
$q^2 > 0$  information about the electromagnetic form factors
can be obtained directly only at $q^2 > 4m^2$ (e.g., from proton-antiproton
annihilation into an electron-positron pair or the inverse reaction),
while the region $0< q^2 < 4m^2$ remains unaccessible in an on-mass shell process.
This is just the region which includes  vector meson production
and thus, could provide some tests of the validity of the vector meson
dominance (VMD) \cite{sacurai} model for the electromagnetic
coupling to  off-mass shell nucleons.

Various models have been elaborated
to calculate the form factors in the time like region
below the $N\bar N$ threshold (see, e.g.,
\cite{wtTjon, vmd_ff,vmd_ff1}) which basically treat the electromagnetic
vertex within an effective meson-nucleon theory in terms
of photon couplings directly to bare nucleons superimposed  to the
coupling to
the meson cloud surrounding the nucleon. Besides, one
can  apply  an analytical continuation of form factors based on
VMD \cite{sacurai} which suggests that the photon
first converts into a vector meson which then couples to hadrons.
This model provides a successful description of the
on-mass shell pion form factor. For nucleons, VMD predicts a strong resonance behavior
of the time like form factors in the neighborhood of  vector meson pole masses.
However, the dipole like behavior of the space like nucleon
form factor persuades us that VMD is too  strong a restriction.
In principle, the original  conjecture of VMD  could be augmented by
introducing heavier  vector mesons (see, e.g., \cite{hohler})
into the parametrization of the nucleon form factor.

The eight independent form factors can be defined in terms of
positive and negative energy projection operators as \cite{wtTjon}
\be
\Gamma_\mu\left(P',P\right)=e\sum\limits_{\rho_1=\pm,\ \rho_2=\pm}
\Lambda^{\rho_1}\left[ F_1^{\rho_1,\rho_2} \gamma_mu
+\frac{i\sigma_{\mu\nu}q^\nu}{2m} F_2^{\rho_1,\rho_2}+
q_\mu F_3^{\rho_1,\rho_2}\right]\Lambda^{\rho_2},
\label{ffs}
\ee
where the third form factor $F_3^{\rho_1,\rho_2}$ is not an independent one,
but is connected with $F_{12}^{\rho_1,\rho_2}$ via the WT identity.
The positive ($\rho=+$) and negative ($\rho=-$) energy
projection operators are denoted as $\Lambda^{\pm}$ respectively. For
the half off-mass shell nucleons only four terms
contribute to the electromagnetic vertex. As mentioned above,
in the time like region below the $N\bar N$ threshold these form factors are to be computed
as loop corrections to the bare electromagnetic vertex \cite{wtTjon}
and/or as analytical continuation of the VMD prediction. Microscopical calculations
\cite{vmd_ff} show that the electromagnetic form factors
are rather sensitive to model assumptions in the region of vector meson pole masses.
At low energies they depend weakly  on  $q^2$
and can be effectively absorbed into the effective parameters of the
$NN\gamma$ Lagrangian \cite{mosel_calc,pascScholten}.

In the present paper we are primarily interested
in di-electron production at intermediate energies with the mass distributions
sufficiently far from the vector meson pole masses so that, following 
\cite{mosel_calc,pascScholten},
we merely put $F_1=1$ and instead of $F_2$ we use the anomalous magnetic
moment of the corresponding nucleon, i.e., we use the Lagrangian (\ref{elm}).
However, for purely  methodological sakes, we also present some results
on di-electron production at
higher energies, where  the $e^+e^-$ invariant mass
covers the $\rho$ and $\omega$ pole masses to evidence effects of the
present VMD implementation.

\subsection{\mbox{\boldmath $\Delta$} isobar}

Intermediate baryon resonances play an important role 
in di-electron production in $NN$ collisions
\cite{fuchs,ernst,nakayama,lutzNew,kapusta,mosel_calc,ikh_model,schafer,brat}.
At intermediate energies the main contribution
to the cross section stems  from the $\Delta$ isobar \cite{mosel_calc}.
Since the isospin of the $\Delta$ is $3/2$
only the isovector mesons $\pi$ and $\rho$ couple to nucleons and $\Delta$.
The form of the effective $\Delta N$ interaction was thoroughly
investigated in literature in connection with $NN$
scattering \cite{bonncd,holinde,weise},
pion photo- and electroproduction \cite{pascQuant,photo,davidson,elctro,lenske}.
The effective Lagrangians of the $N \Delta M$ interactions
read \cite{holinde,weise,pascQuant})
\be
{\cal L}_{\Delta N\pi} = \frac{f_{\Delta N\pi}}{\mu_\pi}
\left[\bar\Psi_\Delta^\alpha \ {\bf T} \ \partial_\alpha \boldPhi_\pi N
\right] +h.c., \label{pionDelta}\\
&& {\cal L}_{\Delta N\rho} =\frac{if_{\Delta N\rho}}{\mu_\rho}
\left[\bar\Psi_{\Delta\,\alpha} {\bf T} \left\{ \partial^\beta\boldPhi_\rho^\alpha-
\partial^\alpha\boldPhi_\rho^\beta\right\}\gamma_\beta\gamma_5 N\right]
+h.c.
\label{rhoDelta}
\ee
with $f_{\Delta N\pi}=2.13 \ GeV$ and $f_{\Delta N\rho}=7.14\ GeV$ \cite{schafer}.
The couplings are dressed by cut-off form factors
\be
F^{ N\Delta M} = \left[ \frac{\Lambda_{ N\Delta M}^2-\mu_M^2 }
{\Lambda_{N \Delta M}^2-k^2}\right]^2,
\label{cutDelta}
\ee
where $\Lambda_{ N\Delta \pi}=1.421\ GeV$ and
$ \Lambda_{ N\Delta \rho}=2.273\ GeV$ \cite{schafer}.
The symbol ${\bf T}$ stands for the isospin transition matrix
(see Appendix~\ref{isospin}), $\Psi_\Delta$ denotes  the field describing the
$\Delta$.  Usually   particles with  higher spins ($s>1$)
are treated within the Rarita-Schwinger formalism in accordance with which
the $\Delta$ field
is a rank-1 tensor (obeying the Klein-Gordon equation)
each component of which is a  4-spinor satisfying the Dirac equation as well.
To reduce the number of redundant degrees of freedom the Rarita-Schwinger
field satisfies also  a number of additional subsidiary conditions
(cf.\ \cite{fronsdal}).
Nevertheless, such a field does not uniquely determine the properties
of  spin-$3/2$ particles.
It is known that an arbitrary field of rank 1 provides a basis for a reducible
representation of the Lorentz group, which can be decomposed into two
irreducible representations corresponding to spins $s=1$ and $s=0$
of the vector field. Correspondingly, an arbitrary solution of field equations
for $\Psi_\Delta$ will be related to spins $s=3/2$ and $s=1/2$.
In order to eliminate the part corresponding to $s=1/2$ one usually
considers spin projection operators \cite{fronsdal}  acting on
an arbitrary solution of the field equations which ensure
the uniqueness of the description of particles with high spins via
\be
\Psi^\alpha_\Delta = P_{\frac32}^{\alpha\beta} \ \Psi^\beta,
\ee
where $\Psi^\beta$ is a solution of the spin-$\frac32$ field
equations and the spin projection operator is defined as
\be
P_{\frac32}^{\alpha\beta}(p) =g^{\alpha\beta} -\frac13 \gamma^\alpha\gamma^\beta
-\frac{2}{3m_\Delta^2} p^{\alpha}p^\beta - \frac{1}{3m_\Delta}
\left( \gamma^\alpha p^\beta - \gamma^\beta p^\alpha\right).
\label{spinproj}
\ee
Then the propagator for the high-spin particles is constructed in
a fully analogous way with the case $s=1/2$. Remaind that formally
the propagator of a Dirac particle with $s=1/2$ can be expressed
as a product of a scalar propagator multiplied
by the positive energy projection operator $\Lambda^+=\hat p + m$ evaluated at
$p^2\neq m^2$. For the Rarita-Schwinger propagator, in order to ensure the propagation
of degrees of freedom  with $s=3/2$, one usually includes  
also the spin projection operator $P_{\frac32}$ to obtain
\be
S_\Delta^{\alpha\beta}(p) =
-\frac{i\left(\hat p + m_\Delta\right)}{p^2 - m_\Delta^2}  \ P_{\frac32}^{\alpha\beta}(p).
\label{raritaprop}
\ee
This propagator is discussed in the literature \cite{will,post,titovprop,tjonprop}
with respect to the fact that for free particles
the positive energy projection operator $\Lambda^+(p)$ commutes with the spin
projection operator $P_{\frac32}^{\alpha\beta}(P)$, which is not the case for
off-mass shell operators.  The adopted
order of their multiplication is  rather a convention than a rule.
Another source of ambiguity is the convention of what to use in (\ref{raritaprop})
as the "particle mass", the on-mass shell value $m_\Delta$ or
the off-mass shell invariant mass $\sqrt{p^2}$ \cite{tjonprop,pascScholten}. Note
that different prescriptions for the propagator differ by  corrections of the
order $p^2-m_\Delta^2$ which, at intermediate energies, could be
absorbed in slight readjustments of the effective parameters.
Indeed, changing the order of the operators in (\ref{raritaprop})
we find an almost constant modification of the cross sections over a wide
range of kinematic variables.

In our calculations we adopted the prescription of \cite{fronsdal,mosel_calc}, i.e.,
the propagator is taken according to Eq.~(\ref{raritaprop}).
In addition, to take into account the finite life time
of $\Delta$, in the denominator
of the scalar part of the propagator, the mass is modified by adding
the  width, i.e., $ m_\Delta \to m_\Delta - i\Gamma_\Delta/2$.
For the kinematics considered here the mass of the
intermediate $\Delta$ can be rather far from its pole value, so that the
width, as a function of $p^2_\Delta$, is calculated
as a sum of partial widths  through the one-pion ($\Delta\to N \pi$)
and two-pion ($\Delta\to N \rho\to N2\pi$) decay
channels \cite{shyamwidts}.

The general form of the $\Delta N \gamma $ coupling
satisfying the gauge invariance can be written
as \cite{davidson,feuster,pascTjon,pascScholten}
\begin{eqnarray}
{\cal L}_{\Delta N \gamma}&=& -i\frac{eg_1}{2m}
\bar\Psi^\alpha \Theta_{\alpha\mu}(z_1)\gamma_\nu
\gamma_5 {\bf T}_3
\psi {\cal F}^{\nu\mu}-  %\nonumber \\ &&
\frac{eg_2}{4m^2}\bar\Psi^\alpha \Theta_{\alpha\mu}(z_2)\gamma_5
{\bf T}_3 \left( \partial_\nu \psi\right) {\cal F}^{\nu\mu} \nonumber\\
&& -\frac{eg_3}{4m^2}\bar\Psi^\alpha \Theta_{\alpha\mu}(z_3)\gamma_5
{\bf T}_3  \psi  \partial_\nu {\cal F}^{\nu\mu} +h.c., \label{deltaLag}\\
\Theta_{\alpha\mu}(z) &=& g_{\alpha\mu}+
[z +\frac12( 1+4z)A]\gamma_\alpha\gamma_\mu,
\label{theta}
\end{eqnarray}
where $A$ is a constant reflecting the invariance of the free $\Delta$
Lagrangian with respect to point transformations \cite{rarita}. Since observables
must not depend on this parameter, $A$ is  arbitrary. According to common
practice one puts $A=-1$. The other parameter, $z$, is also connected
with point transformations, however it is a characteristic of the off-mass shell
$\Delta$ resonance and remains unconstrained. The meaning of this parameter
is that every coupling to a  spin-$3/2$ field contains also contributions from couplings
with spin-$1/2$ components. Some times $z$
is called  the off-mass shell parameter. Investigations of the role  of
the off-mass shell quantity, treated as a free parameter,
in different observables related to the  $\Delta$ show \cite{feuster,davidson} that
inclusion of $z$ into the calculations   requires a
slight readjustment of the effective parameters $g_i$, which
are also free parameters. This means that the parameter $z$ and the effective couplings
$g_i$ must be simultaneously adjusted to given observables.
A thorough study of the role of couplings to spin-$1/2$
particles \cite{pascQuant,pascScholten}
has shown that the dependence on $z$  is rather weak, making the off-mass shell parameter
redundant (see also discussion in \cite{mosel_calc}).
Basing on this observation, we  neglect the off-mass shell parameter by merely putting
$\Theta_{\alpha\mu}(z) = g_{\alpha\mu}$. The coupling constants $g_i$
are taken as in \cite{mosel_calc}, i.e.,
$g_1=5.416$, $g_2=6.612$ and $g_3=7.0$.

\subsection{Results for \mbox{\boldmath $NN \to NN$}} \label{subsecTri}

As mentioned  above our effective parameters have been fixed in such a way as
to reproduce reasonably well the results of the study~\cite{mosel_calc} performed to explain
the  DLS data \cite{DLS} at low energies.
The OBE parameters (listed in Table \ref{tabl1}) and their energy dependence
have been taken as in Ref.~\cite{mosel_calc}. Figures~\ref{fig4} and \ref{fig5}
show results of our calculations of the  mass distribution of di-electrons
in $pp$ and $pn$ collisions at two values of the kinetic energy,
$1.04\ GeV$ and $2.09\ GeV$, corresponding to those considered in \cite{mosel_calc}.
The dotted lines depict the contribution  of pure bremsstrahlung
processes from  nucleon lines, i.e., di-electrons are produced solely due to
nucleon-nucleon interaction via the one-boson-exchange potential. In our actual calculations
we include four exchange mesons, $\pi$, $\sigma$, $\rho$ and $\omega$ mesons 
% (cf.\ Fig.~\ref{fig2}), 
supplemented by a "counter" term simulating  a heavy axial vector-isovector meson,
with the goal  to cancel singularities
of the pion potential at the origin \cite{mosel_calc}.

The dashed lines in Figs.~\ref{fig4} and \ref{fig5} depict the
contributions of the $\Delta$ isobar within the same OBE potential.
%Note that only charged mesons, $\pi$ and $\rho$, contribute to such diagrams. 
The solid lines represent the total cross section including all interferences.
It can be seen that for $pp$ collisions almost in the whole kinematical range 
the $\Delta$  contribution dominates. 
Near the  kinematical limits the nucleon contribution becomes
comparable with $\Delta$ contributions. At smaller invariant masses, the $N$
and $\Delta$ contributions interfere constructively, while near the kinematical
limit and at beam energy $2.09\: GeV$ the interference pattern
in $pp$ reactions is just so that the total cross section
resembles the $N$ or $\Delta$ contributions individually.
Another observation is  the fact
that, due to isospin factors the cross section in $pn$ reactions 
is systematically larger than in $pp$ reactions. 
These results are in agreement with \cite{mosel_calc}.
In $pn$ collisions the inclusion of the contact terms amplifies the 
contribution of pure nucleon diagrams. 

After adjusting the model parameters we proceed and present
in Figs.~\ref{fig6} and \ref{fig7} results at energies envisaged in the approved
HADES proposal \cite{HADES} for $pp$ and $pn$ processes.
It is seen that, except for the absolute values, the
behavior of the cross section and the relative contributions
of $\Delta$ isobars and pure nucleon bremsstrahlung basically does not change with energy.
However, as seen from these figures, the kinematical range of the di-electron mass
becomes essentially larger covering also the region of
vector  meson production, i.e., $\rho$ and $\omega$, which  have
been not yet implemented
in the calculations. Consequently, at these energies the results presented
in Figs.~\ref{fig6} and \ref{fig7} are to be considered as an estimate
of a smooth bremsstrahlung background. Effects of $\rho$ and  $\omega$ excitations will be
considered in the next subsection.

In Fig.~\ref{fig8} the isospin effects in $pn$ and $pp$ reactions are quantified. The
dot-dashed line illustrates the difference between $pp$ and
$pn$ processes in pure nucleon bremsstrahlung, the dashed line reflects the isospin effects
for the $\Delta$ contribution, while the full line is the ratio of the total cross sections.
Note the nontrivial invariant mass dependence which prohibits the use of simple,
constant isospin factors to relate $pp\to pp\: e^+e^-$
and  $pn\to pn\: e^+e^-$ cross sections.

\subsection{ VMD effects} \label{subsecVMD}

As seen in Figs.~\ref{fig6} and \ref{fig7}, at beam energies $T_{kin} > 2\: GeV$ 
the kinematical range of
the invariant mass $M$ covers the $\rho$ and $\omega$ pole masses
(corresponding to $0.768$ and $0.783\ GeV$, respectively)
so that in this region the vector meson nature of the electromagnetic coupling of photons
with nucleons can show up. In simple terms, the  VMD hypothesis \cite{sacurai}
implies that photons couple to hadrons ($\pi$ mesons, nucleons etc.)
solely via  intermediate vector mesons, in which case the electromagnetic
form factor reads
\be
F_{VMD}(q^2)=\frac{M_V^2}{M_V^2-q^2}.
\label{VMDff}
\ee
Such a behavior of form factors
can be obtained in a more rigorous way within an effective meson-nucleon theory, 
like the one used in the present
paper. For this purpose one should  consider additionally effective Lagrangians 
with electromagnetic
couplings of the vector mesons (only $\rho$ and $\omega$ in the
kinematical region we are interested in)  with photons. This procedure
is not unique and one should pay attention  to avoid double counting of contributions
from the Lagrangian
with  direct $NN \gamma$ coupling. Usually \cite{lutzNew,vmd_ff,vmd_ff1,friman1}
the electromagnetic $\gamma\rho$ and $\gamma\omega$ interaction
Lagrangians are added to the $NN \gamma$ Lagrangian (\ref{elm}) and are chosen in the form
\be
{\cal L}^{em}_{\rho(\omega)\gamma}
=-\frac{e}{2f_{\rho(\omega)\gamma}}{\cal F}^{\nu\mu}{\cal G^{\rho(\omega)}}_{\nu\mu},
\label{rhogamma}
\ee
where
${\cal G^{\rho(\omega)}}_{\nu\mu}$ is the field strength tensor of the $\rho\: 
(\omega)$ meson.
Note that the Lagrangian (\ref{rhogamma}) should be considered only together with the
Lagrangians (\ref{mnn}) and  (\ref{elm}), in which case the
proton electromagnetic vertex reads
\be
\Gamma^\mu_{NN\gamma} = \left( \gamma^\mu -\displaystyle\frac{q^\mu \hat q}{q^2}\right )F_1(q^2)
+\displaystyle\frac{q^\mu \hat q}{q^2} +
i\kappa F_2(q^2) \frac{\sigma^{\mu\nu}q_{\nu}}{2m},
\label{newff}
\ee
where
\begin{eqnarray} &&
F_1(q^2)=1 + \frac{g_{ NN\rho}}{f_{\rho \gamma}}
\left [  \frac{q^2}{M_{\rho}^2-q^2}+\frac{f_{\rho\gamma}}{f_{\omega\gamma}}
\frac{g_{ NN\omega}}{g_{ NN\rho}}  \frac{q^2}{M_{\omega}^2-q^2}  \right],\nonumber\\&&
F_2(q^2)=1+\frac{g_{ NN\rho}}{f_{\rho \gamma}}
\frac{q^2}{M_{\rho}^2-q^2}.
\label{f1VMD}
\end{eqnarray}
Note that the vertex function (\ref{newff}) obeys  the WT identity.
The introduced coupling constants can be estimated
from VMD like pole fits \cite{hohler,dubnicka,gari} and also from the requirement
that at $q_\mu\to 0$ one has $F_1\to (1+\tau_3)/2$
(see also \cite{wtTjon,friman1}). As a result, one can
approximately take
\be
\frac{g_{ NN\rho}}{f_{\rho \gamma}}=\frac12,\quad
\frac{f_{\rho\gamma}}{f_{\omega\gamma}}\frac{g_{ NN\omega}}{g_{ NN\rho}}=1,
\label{edin}
\ee
which provides  the following form of  the form factors
\begin{eqnarray} &&
F_1(q^2)=1 + \frac12
\left [  \frac{q^2}{M_{\rho}^2-q^2}+ \frac{q^2}{M_{\omega}^2-q^2}  \right],
\label{pervyi}\\&&
F_2(q^2)=1+\frac12
\frac{q^2}{M_{\rho}^2-q^2}.
\label{FFVMD}
\end{eqnarray}
Since both $\rho$ and $\omega$ are not stable the corresponding masses in
(\ref{pervyi}) and (\ref{FFVMD}) receive also imaginary parts, 
i.e., $M_V\to M_V -iM_V \Gamma_V/2$,
where $\Gamma_V$ is the total decay width of the respective vector meson. In our calculations
we take advantage of the fact that a free $\rho$ meson decays
mainly into two pions, so that
its width, as a function of the invariant mass $q^2$,  
can be calculated within the same effective
meson-nucleon theory with the result
\be
\Gamma_\rho (q^2)=\Gamma_\rho(M_\rho^2)\frac{M_\rho^2}{q^2}\left(
 \displaystyle\frac{\sqrt{q^2-4\mu_\pi^2}}{\sqrt{M_\rho^2-4\mu_\pi^2}}\right)^{3},
 \ee
where $\Gamma_\rho(q^2=M_\rho^2) \approx 0.15\  GeV$. 
The width of the $\omega$ meson in the present calculations 
has been kept constant $\Gamma_\omega = 0.03\ GeV$
to simulate the finite resolution \cite{HADES}
(HADES envisages an invariant mass resolution $\Delta M/M\sim 1\%$ \cite{holzman}).
Note that the VMD model can be recovered if,
as usually, one takes $M_\rho\approx M_\omega=M_V$ and $\Gamma_\rho\approx\Gamma_\omega$.

Some comments are in order here.
The VMD hypothesis could be implemented not only via the effective Lagrangians
(\ref{rhogamma}), (\ref{mnn}) and (\ref{elm}) but also by considering the simplest form for the
$\gamma V$ coupling
\cite{wtTjon,kroll,friman2}
\be
{\cal L}_{VMD}=-\frac{e M_{\rho(\omega)}^2}{f_{\rho(\omega)\gamma}} 
\Phi^\mu_{\rho(\omega)} A_\mu
\label{tjonVMD}
\ee
{\it without} the direct term (\ref{elm}). The Lagrangian (\ref{tjonVMD}) corresponds  better
to the original VMD conjecture \cite{sacurai} since it assumes that the
electromagnetic coupling
$ NN\gamma$ occurs solely via the vector mesons.  It is easily  seen
that if one employs the Lagrangians (\ref{mnn}) and (\ref{tjonVMD}),
e.g., the form factor $F_1$ with $M_\rho\approx M_\omega=M_V$ and
$\Gamma_\rho\approx\Gamma_\omega$,
the VMD form (\ref{pervyi}) or (\ref{VMDff})  could be  obtained \cite{wtTjon}.
An inclusion of direct terms, like Eq.~(\ref{elm}),
will lead to double counting in the corresponding amplitude. Hence, the Lagrangian
(\ref{tjonVMD}) simultaneously accounts for  vector meson production effects near the
corresponding pole masses ($M_\rho$ and $M_\omega$) and for the background (direct)
contribution in the whole kinematical range. Consequently a separation of these two
kinds of effects is hampered with the VMD Lagrangian taken as in Eq.~(\ref{tjonVMD}).
In our calculations we use the Lagrangians (\ref{elm}) and (\ref{rhogamma})
which allow to distinguish the contribution of direct terms  from the
vector meson production, i.e., the first term in Eqs.~(\ref{pervyi}) and
(\ref{FFVMD}) is referred to as the direct or background contribution,
while the second one defines the $\rho$ and $\omega$ meson contribution
(see also \cite{vmd_ff}).

Figures \ref{fig9} and \ref{fig10} illustrate the effects of
VMD at two kinetic beam energies, $T_{kin}=2.2 \ GeV $ and $T_{kin}=3.5\ GeV $,
for $pp$ and $pn$ collisions, respectively. In the  upper panels we present VMD effects
for the pure nucleon contribution, while in the lower panels 
the $\Delta$ contribution is included
as well. The dashed lines represent the background cross section, i.e., the one calculated
with only the first term in Eqs.~(\ref{pervyi}) and (\ref{FFVMD}) 
(cf.\ Figs.~\ref{fig6} and \ref{fig7}).
The dotted lines have been obtained when only the second ($\rho$) or the
third terms ($\omega$) have been
taken into account. It is seen that the resonance structure of the $\rho$ contribution
is not so pronounced  being rather broad, because of
its relatively low threshold and because of the mass dependence of its the decay width.
Contrarily, the $\omega$ contribution has a rather sharp, resonance like behavior.
The relative contribution of $\omega$ and $\rho$ near the pole mass,
$M_0\sim M_\omega\sim M_\rho\sim 0.78\ GeV$, is
basically governed by the ratio of meson widths squared,
$\Gamma_\omega^2(M_0)/\Gamma_\rho^2(M_0)=0.03^2/0.15^2$
(remind that we attribute to the $\omega$ meson the actual width of $30\: MeV$
to simulate finite  detector resolution).

Here it is worth stressing that  within the VMD model the vector
mesons are designed to mediate the electromagnetic
coupling of photons  with nucleons. Consequently,
they contribute in the whole kinematical range of the invariant mass
and, except the neighborhood of the pole masses,
are essentially virtual. This implies that, apart from the intervals near the pole masses,
the cross section can not be presented as a two-step process consisting of:
(i) production of a vector meson resonance with experimentally known width and
with a mass  around the $\rho$ and/or $\omega$ mass, (ii) the  independent
subsequent decay
into a di-electron channel (cf. discussion in Ref.~\cite{knoll}).
The di-electron emission within
the VMD model  is rather a process of production of a {\it virtual}
vector particle with the quantum numbers of the $\rho$ or/and $\omega$
with subsequent decay into a di-electron (via an intermediate
conversion into a virtual photon) which contributes in the
whole kinematical range of the di-electron invariant masses. This issue is discussed
in some detail in Appendix \ref{ts}.

\subsection{Effects of final state interaction }
Previous studies of threshold-near vector meson production in $NN$ reactions
have shown \cite{nakayama,ourOmega,titovprop,titov} that
the final state interaction (FSI)
between nucleons plays an important role. It has been also
found that considerable corrections from FSI
occur at low values of the energy excess, $\Delta s^{\frac12} = \sqrt{s} -2m -M_V$,
where the relative momentum (excitation energy) of
the nucleon pair is small. With increasing energy excess the relative momentum increases
too and FSI effects become less important \cite{nakayama,ourOmega}.

 In reactions of di-electron production the kinematical situation is
 rather different. As seen from Eq.~(\ref{diffcross})  the invariant  mass, $s_\gamma$,
 of the $e^+e^-$ pair  varies from the photon point  to a maximum
 value dictated by  kinematics. Similar to  the case of on-mass shell vector
 meson production (cf.\ Eq.~(2.4) in Ref.~\cite{ourOmega}),
 in  Eq.~(\ref{diffcross}) an integration  over
 the excitation energy  $s_{12}$ is to be performed. However, in this case the
 kinematical range of $s_{12}$ and, consequently,
 the range of the relative momentum of the nucleon pair is rather
 different and strongly depends on the value of the di-electron  invariant  mass.
% To find a correspondence with previous results~\cite{nakayama,ourOmega}
% let us  define an effective excess energy  which has
% the same  meaning as  for the
% reaction $N_1+N_2\to N_1'+N_2'+V$~\cite{nakayama,ourOmega},
% but now depending on the di-electron mass,
% \begin{equation}
% \Delta s^{\frac12}_{eff} = \sqrt{s} -2m -\sqrt{s_\gamma}.
%\label{effectives}
%\end{equation}
%It is clear that the magnitude of $\Delta s^{\frac12}_{eff}$ kinematically
%governs the magnitude of the relative momentum of the final nucleons. Hence
%the FSI effects in di-electron production are expected to be of the same order of magnitude as in
%on-mass shell processes evaluated at equal $\Delta s^{\frac12}_{eff}$ and $\Delta s^{\frac12}$,
%respectively. So, at kinetic energy $T_{kin}=2.2\: GeV$ the energy excess in on-mass shell
%reactions is $\Delta s^{\frac12}\simeq 0.1\: GeV$, while $\Delta s^{\frac12}_{eff}$
%varies from zero (at $s_\gamma=s_{\gamma\: max}=(\sqrt{s}-2m)^2$)  to $\Delta s^{\frac12}_{eff}\simeq 0.9 \: GeV $
%(at $ s_\gamma=s_{\gamma\: min}=0$). This means that
With  increasing di-electron mass  FSI effects are expected
to increase. For instance, the results of  Ref.~\cite{ourOmega}
indicate  that for $pn$ reactions
at di-electron mass
near  the $\omega$ pole mass, FSI effects
lead to an increase of the cross section by a factor $\sim 2$, while
at lower masses the FSI effects are expected to diminish.
However, with increasing  di-electron  mass, the kinematical
range of $s_{12}$ shrinks, so that FSI effects are expected to increase.

To take into account FSI we employ here the same model as in
a previous study \cite{ourOmega} of the vector meson production
based on the Jost function formalism \cite{gillespe}, which provides a good
description of $NN$ interaction and phase shifts at low relative momenta.
Detailed results are displayed in Fig.~\ref{fig10a}, where FSI
effects have been  calculated at four values of kinetic beam energy.
The solid lines illustrate the magnitude of FSI corrections, while the
dashed lines correspond to results obtained without taking
into account FSI (cf.\ Figs.~\ref{fig7} and ref{fig10}).
As expected, FSI plays a minor role at low values of the
di-electron invariant mass but increases at the kinematical limit. 
For completeness, in the lower
row we present also results with VMD included, where  dotted (dot-dashed)
lines denote  the cross section
without VMD, including (excluding) FSI.
From this figure we conclude, that throughout  the kinematical range of interest,
effects of FSI are not too large ($20 - 50 \: \%$), but become essential and even
dominant at the kinematical limits.
The cross section exhibits a similar behavior in $pp$ reactions and,
therefore, are not exhibited here.

\section{Di-electrons in the process $\mathbf { D p \to p_{\rm sp} p   n  e^+ e^-}$}\label{Deu}

Now we are going to implement the parametrization of the amplitude of the
process $n p \to n p  e^+ e^-$ in the exclusive reaction  $D p \to p_{\rm sp} p   n  e^+ e^-$.
The latter process will be studied by the HADES collaboration \cite{HADES,holzman}
with the above mentioned goals.

\subsection{Formalism}\label{subDeu1}

Let us  consider the reaction
\begin{equation}
D (P_D) + p(P_1) = p(p_{\rm sp}) +n(P_1)'+p(P_2')+e^+(k_1)+e^-(k_2)
\label{reactiond}
\end{equation}
within the spectator mechanism. The internal neutron of the deuteron
interacts with the target proton producing a di-electron as a consequence
of bremsstrahlung processes, while the detected (forward) proton
acts as a spectator. In principle, there could be di-electron emission from
the detected proton, due to final state interaction effects in the three nucleon system.
The FSI  of the spectator with  the active $pn$ system can be estimated
within a generalized eikonal approximation  (see, e.g.\ \cite{fs}) by considering
the electro-disintegration of the $^3He$ in processes $^3He(e,e'p)pn$. A detailed study of such
processes \cite{braunFSI} shows that in  parallel kinematics the FSI effects can be safely
disregarded and, consequently, the di-electron production
off the spectator can be neglected indeed.
In the present paper the parallel kinematics is guaranteed
by the choice of the direction of the
detected proton in the very forward direction,
say at $0^\circ -5^\circ$, as envisaged in the experimental proposal
\cite{HADES}. The FSI
effects in the active $pn$ pair depend on the di-electron invariant mass.
Near the  kinematical limit the relative momentum of the
$pn$ pair becomes small, therefore an enhancement of the 
FSI effects is expected in this region.

As in the previous section, we choose the kinematics with two intermediate invariant masses
$s_\gamma$ and $s_{12}$, as depicted in Fig.~\ref{fig9}.
The invariant differential cross section reads
\begin{eqnarray}
d\sigma  &=&
\frac{1}{2(2\pi)^{11}\sqrt{\lambda(s,M_D^2,m^2)}}\,\frac{1}{6}
\sum\limits_{{\cal M}_D,spins}
\, |T|^2 \frac{ d^3p_{\rm sp}}{2E_{\rm sp}}\
d s_\gamma\ ds_{12}\\
&\times&
R_2 \left( P_1+P_n\to P_\gamma+P_{12}\right)\
R_2 \left( P_\gamma\to k_1+k_2 \right)\
R_2 \left( P_{12}\to P_1'+P_2'\right), \nonumber
\label{equ}
\end{eqnarray}
where $M_D$ is the deuteron mass.
Integration over lepton variables can be performed as above.
The same effective meson nucleon theory as in the previous section is employed
to parameterize $\vert T \vert^2$.
In order to keep the covariance of the formalism
and to use directly all the previous results we compute the corresponding
hadronic electromagnetic current within the Bethe-Salpeter (BS) formalism.
The current is now
\begin{eqnarray}
J_\mu =\bar u({\bf P}'_1,s_1')\bar u\left({\bf P}_2',s_2'\right) {\cal O}_\mu
u\left({\bf P}_1,s_1\right) \Psi_D^{\CM_D}\left(P_n,p_{\rm sp}\right)
\widetilde S^{-1}(p_{\rm sp}) v\left({\bf p}_{\rm sp},s_{\rm sp}\right),
\label{deucurr}
\end{eqnarray}
where $u\left({\bf P},s\right)$ and $ v\left({\bf P},s\right)$
are free Dirac spinors. The operator ${\cal O}_\mu$ is a short-hand notation
for the operators of di-electron production in $NN$ interactions  within
the adopted approach. Actually, ${\cal O}_\mu$ represents the set of diagrams
depicted in Figs.~\ref{fig2} and \ref{fig3} with all nucleon and photon lines truncated;
the BS amplitude for the deuteron
with total spin projection $\CM_D$ is denoted as $\Psi_D^{\CM_D}$, and the
(modified) inverse propagator of the spectator is
$\tilde S^{-1}(p_{\rm sp})\equiv (\hat p_{\rm sp}+m)$.

Since our numerical solution for the BS equation has been obtained in the
deuteron center of mass \cite{solution}, all further calculations will be performed in
this system, i.e., in the "anti-laboratory"  system.
In general, the BS amplitude consists of eight partial components.
We take into account here the most important ones, namely the
$S$ and $D$ partial amplitudes. The other six amplitudes may become important
only at high transferred momenta \cite{ourphysrev,quad}, hence for the present
process (\ref{reactiond}) with forward detection of the spectator,
they may be safely disregarded. Observe that the quantity
\be
{\cal V}_D\equiv \Psi_D^{\cal M}\left(P_n,p_{\rm sp}\right)
\widetilde S^{-1}(p_{\rm sp}) v\left({\bf p}_{\rm sp},s_{\rm sp}\right)
\label{vd}
\ee
being a four dimensional column in the spinor space,
acts as a "deuteron spinor" and formally replaces in  the deuteron current $J_\mu$
the corresponding neutron spinor.

The square of $J_\mu$ is given by
\begin{eqnarray}
\frac16 \sum\limits_{\CM, spins}\ J_\mu\ J^{+\mu}=\frac16 \sum\limits_{\CM, spins}
Tr\left[ \left(\hat P_2'+m\right ){\cal O}_\mu
{\cal V}_D\bar {\cal V}_D {\cal O}^\mu\right].
\end{eqnarray}
A direct evaluation (see Appendix \ref{a2}) yields
\be
\frac16 \sum\limits_{\CM, spins}
{\cal V}_D\bar {\cal V}_D=2M_D (2\pi)^3 n_D\left (|{\bf p}_{\rm sp}|\right)
\frac14 \left( \hat k_n+m\right) \label{vv},
\ee
so that
\begin{eqnarray}
\frac16 \sum\limits_{\CM, spins} J_\mu J^{+\mu} &=&
2M_D (2\pi)^3 n_D\left (|{\bf p}_{\rm sp}|\right)\frac14
Tr\left[ \left(\hat P_2'+m\right ){\cal O}_\mu
\left(\hat k_n+m\right ) {\cal O}^\mu\right]\nonumber\\
&=&
2M_D (2\pi)^3 n_D\left (|{\bf p}_{\rm sp}|\right)\frac14\sum_{spins}
\left( J_\mu\ J^{\mu +}\right)_{NN},
\label{rel}
\ee
where the deuteron momentum distribution is
\be
n_D\left (|{\bf p}_{\rm sp }|\right)=\frac{1}{2\pi^2}
\left (U_S \left (|{\bf p}_{\rm sp}|\right)^2+
U_D\left (|{\bf p}_{\rm sp}|\right)^2\right)
\label{deumomd}
\ee
with normalization
$\int n_D\left (|{\bf p}_{\rm sp}|\right) d^3 {\bf p}_{\rm sp} \approx 1$.
(Note that the normalization is not exactly
unity because, within the BS formalism, besides the main $S$ and $D$ partial waves
other components, e.g., the negative $P$ waves,
enter in the definition of  the total momentum distribution. Their contribution is
however extremely small \cite{quad,Tjon}.)

Equation (\ref{rel}) relates the deuteron cross section with the subprocess
of di-electron production in proton-neutron interactions via
\be
2E_{\rm sp}\frac{d\sigma}{d^3{\bf p}_{\rm sp}d s_\gamma}=2M_D\sqrt{\displaystyle
\frac{\lambda(s_{NN},m^2,m^2))}{\lambda(s_0,m^2,M_D^2)}}
n_D\left (|{\bf p}_{\rm sp}|\right)
\displaystyle\frac{d\sigma^{np}}{d s_\gamma}.
\label{crossD}
\ee
Remarkable is the factorization of the cross section, for which the needed
formulae are outlined in Appendix \ref{a2}.

\subsection{Results }\label{subDeu2}

We have calculated the di-electron production in the exclusive process (\ref{reactiond})
at three values of the kinetic energy envisaged at HADES~\cite{HADES}, $T_{kin}=1.25 $,
$1.90 $ and $ 3.5\ A\ GeV$. As mentioned above, the effects of the final state interaction
of the spectator nucleon with the "active nucleon"
are minimized within the parallel kinematics,
where the spectator is detected essentially in the same direction 
as the incident deuteron with
approximately the same velocity.
In our actual calculations we specify, at each considered energy, 
three angles for the spectator in the
forward direction, $\theta=1^\circ,\: 2^\circ$ and $\: 5^\circ$. 
The remaining two independent
kinematical variables, the momentum of the spectator $p_{\rm sp}$ 
and the di-electron invariant mass
$M$, are considered in a large kinematical range.

In Fig.~\ref{fig12} the dependence
of the cross section on the spectator momentum is exhibited for two fixed values
of the invariant mass and for $\theta=1^\circ$. The three curves (solid, dashed and dot-dashed)
in each panel correspond to three different beam energies. It is seen that at each
energy the cross section has a maximum at the spectator momentum
$ \bsp =\displaystyle\frac12 \bpD $  (in the anti-laboratory system this corresponds to $\bsp=0$).
The widths of the distributions increase with increasing energy, which
is merely an effect of the larger phase space volume.

In Fig.~\ref{fig13} the mass distribution is depicted.  
The three columns (left, middle and right)
correspond to three different angles ($\theta=1^\circ,\: 2^\circ$ and $5^\circ$),
while the three rows (upper, middle and lower) specify three kinetic beam energies,
$T_{kin}=1.25, \:  1.9,$ and $3.5\ A\: GeV$. For such kinematical conditions the invariant
cross section has been calculated at three different values of the spectator momentum,
$\bsp=0.45\: \bpD$ (dot-dashed curves), $\bsp=0.5\: \bpD$ (solid curves), and
$\bsp=0.55\: \bpD$ (dashed curves), around the maxima of the cross section 
(cf.\ Fig.~\ref{fig12}).
The results in Figs.~\ref{fig12} and \ref{fig13} do not  yet include VMD effects
in the $NN$ subprocess, i.e., the results  can be considered as estimates of the  background
contribution.

Fig.~\ref{fig13} illustrates that the shape of the cross section as a function of the
invariant mass basically reproduces the one in the $pn$ subprocess. It also seen that
at fixed energy and angle the effect of variation of the spectator momentum
(dot-dashed, solid and dashed lines in each panel), apart from decreasing the
subprocess's  phase space volume with increasing momentum, reduces to a factor being
proportional to the  deuteron momentum distribution $n_D\left (|{\bf p}_{\rm sp}|\right)$
(here ${\bf p}_{\rm sp}$ is the spectator momentum in the anti-laboratory system), as seen from
Eq.~(\ref{crossD}). At moderate values of the invariant mass, e.g., not too close to the
kinematical limit, the most favorable conditions for di-electron detection are
low angles $\theta\sim 1-2^\circ$ and $\bsp=\displaystyle\frac12 \bpD$. With $\theta$ increasing
the cross section at lower values of $p_{\rm sp}$ becomes comparable
with  the cross section at  $\bsp=\displaystyle\frac12 \bpD$. This can
be explained that, in spite of  $\bsp < \displaystyle\frac12 \bpD$,
the  deuteron momentum distribution $n_D\left (|{\bf p}_{\rm sp}|\right)$
decreases (in anti-laboratory system  $|{\bf p}_{\rm sp}|\neq 0$ holds), but
the phase space increases faster so that
the total cross section becomes even larger than at $\bsp=\displaystyle\frac12 \bpD$
(where in the anti-laboratory system $|{\bf p}_{\rm sp}|= 0$),
as the right column in Fig.~\ref{fig13} clearly exhibits.

Finally, in Fig.~\ref{fig14} we present results with VMD effects implemented
at the kinetic energy $T_{kin}=3.5\: A\: GeV$, for which the kinematical
range of the invariant mass covers the region of $\rho$ and $\omega$ pole masses. As in the
$pn$ subprocess the cross section sharply increases near the vector meson poles. The
contribution to the peak comes mainly  from  the $\omega$ excitation.
As can be seen in Figs.~\ref{fig7} and \ref{fig12} at  $T_{kin}=1.9\,A\: GeV$
the kinematical limit of the di-electron invariant mass is located just in the vicinity of the
vector meson pole masses. In this region the phase space volume for 
$NN$ reactions shrinks to zero
and all possible effects of VMD are masked. However, in the deuteron case the phase space
volume can be enlarged by considering  spectator momenta 
with velocities smaller than the initial one.
This implies  that in the subsystem of the two active nucleons the total energy is larger
than in the free $NN$ kinematics. Consequently,  
effects of sub-threshold vector meson production
can be observed in this region.
The lower the spectator momentum the larger kinematical range 
of the allowed di-electron invariant mass can be achieved.
However, since the deuteron internal momentum distribution sharply decreases
with  increasing  spectator momentum, the subthreshold di-electron production
at very small $\bsp < \frac12 \bpD$ (in anti-laboratory the spectator is backward with
increasing $|{\bf p}_{\rm sp}|$)
is prohibited. Therefore, it is clear that at threshold-near energies there
should be a restricted interval for  the spectator momentum
within which  an experimental investigation of the
subthreshold production of vector mesons can be achieved. At too large
spectator momenta ($\bsp\gtrsim \displaystyle\frac12 \bpD$) 
the suppression originates from the shrinking
phase space, whereas too small momenta are restricted by the deuteron's internal
momentum distribution.

In Fig.~\ref{fig15} we present our results for
the cross section (\ref{crossD})
at the near-threshold energy $T_{kin}=1.9\ A \: GeV$ at two
values of the spectator momentum detected in the very forward direction, $\theta = 1^\circ$.
For such a forward kinematics it is quite easy to estimate the effects
of enlarging the phase space
for the elementary subsystem. At $T_{kin}=1.9\: A\: GeV$ the deuteron momentum is
$\bpD =5.36 \: GeV/c$ so that, e.g., for  the spectator momentum $\bsp=0.25\:  \bpD$,
the momentum of the active neutron before interaction
is $|{\bf p}_n | \simeq 0.75\:  \bpD \simeq 4.02\: GeV/c$.
This corresponds to
a kinetic energy of $T_{kin}\simeq 3.2\: GeV$ in the $np$ subsystem above the vector
meson production threshold, hence the cross section at low values
of the spectator momenta can leak away into the
kinematically forbidden region for the free $np$ process,
as seen in Fig.~\ref{fig15}.
The solid lines
correspond to the spectator momenta $\bsp=0.25\: \bpD$
(left panel)  and  $\bsp=0.35\: \bpD$ (right panel) respectively.
One observes that
as far as the invariant mass is not too close to the threshold, the
cross section for the quasi-free kinematics is much larger than at
$\bsp \neq \displaystyle\frac12\: \bpD$, i.e., the suppression
caused by  the deuteron momentum
distribution is more important than the effect of enlarging the phase space.
In the region close to threshold the quasi-free cross section falls rapidly
to zero and the subthreshold cross section becomes predominant. It is also
seen that the contribution of $\rho$ and $\omega$ production (dotted lines)
is by one order of magnitude above the background.  In this case the
contribution from the background (the first terms in Eqs.~(\ref{pervyi}) and (\ref{FFVMD}))
can be safely neglected, hence a direct investigation of the vector meson
production becomes feasible.

We also have investigated effects of FSI in the reaction
$Dp\to p_{\rm sp} pn\: e^+e^-$. As we  mainly consider
the parallel kinematics, FSI of the spectator proton with
the active $np$ pair can be safely neglected and  FSI can
be important  only in the active $np$ pair.
In complete agreement with the factorization formula and
previous estimates \cite{ourOmega,titov}
the effects of FSI are small at small values of the di-electron mass but
become visible and important at the
kinematical limit. This is illustrated in Fig.~\ref{fig17},
where calculations are presented for
$T_{kin}=\:  1.9\: A\:  GeV$ for two values of the spectator momenta.
The solid and dashed lines depict results for  the spectator
momentum $\bsp =\frac12\: \bpD $, while the dot-dashed and dotted lines
are for  $\bsp =0.55\: \bpD$. The former value of
$\bsp $ corresponds to the quasi free kinematics, i.e.,
to the energy of the active  $np$ pair near the vector meson production
threshold. The latter one determines subthreshold energies even for the
$np$ subsystem.
Thus, for both values of the spectator momenta the vector meson production threshold
is hardly reached and, consequently, FSI effects here are maximized.
At lower values of the spectator momentum, as well as at lower di-electron
invariant mass, the FSI effects are negligibly small.

\section{Summary} \label{summary}

In summary
we have analyzed different aspects of the di-electron production from the
bremsstrahlung mechanism at energies envisaged at HADES \cite{HADES} for
the exclusive reactions $NN\to NN\: e^+e^-$ and  $Dp\to p_{\rm sp} pn\: e^+e^-$. 
To calculate
the corresponding cross sections we employed an effective meson-nucleon theory with
parameters adjusted to elastic $NN$ and inelastic $NN\to NN\pi$ \cite{mosel_calc}
reaction data with $\Delta$ isobars included and with account of
vector meson dominance effects.
The performed evaluations of bremsstrahlung diagrams can be considered
as an estimate of the background contribution, a detailed knowledge of which is a necessary
prerequisite for understanding di-electron production in heavy-ion collisions.
Our approach is based on covariant evaluations of the corresponding tree level Feynman
diagrams with implementing phenomenological form factors and
vector meson dominance  effects, with particular attention paid 
on preserving the gauge invariance.
The covariance of the approach is achieved by direct relativistic calculations
of  Feynman diagrams for $NN$ collisions and by implementation of the Bethe-Salpeter
formalism for the $Dp$ reaction. The latter case bases on
our previously obtained solution of the homogenous Bethe-Salpeter equation
with realistic interaction \cite{solution}.

In accordance with previous results \cite{mosel_calc} our calculations demonstrate
that in the region of invariant masses far from the vector meson production
threshold the main contribution to the cross section, in both reactions $pp$ and $pn$,
comes from virtual excitations of  $\Delta$ isobars. Due to isospin effects the cross
section for the reaction $pn\to pn\: e^+e^-$  is larger than the cross section 
for  $pp\to pp\ e^+e^-$
by a factor $ 1.5-3$. Note that  because of  
(i) contributions
of the isoscalar $\sigma$ and $\omega$ exchange mesons, 
(ii) differences
in the electromagnetic coupling in $\gamma p$ and $\gamma n$ systems, and 
(iii) interference effects,
the isospin enhancement is not $\sim 9$, as one could naively expect from isospin
symmetry considerations. In both reactions, $pn\to pn\: e^+e^-$ and $pp\to pp\ e^+e^-$,
the bremsstrahlung cross section exhibits a smooth behavior as a function of the
di-electron mass. Hence, the bremsstrahlung cross section
can be considered as background contribution. In the
kinematical range close to the vector meson pole masses the cross section
has sharp maxima, clearly indicating that the di-electron production
can  be also considered as a tool of testing
the validity of vector meson dominance. The performed investigations
of the vector meson dominance effects show that they contribute in the whole kinematical range.
Near the pole masses  our diagrammatical approach  provides a form for
the cross section resembling a two-step model formulae,
however it preserves the possibility to trace back the essential differences between
the two models (see also discussion in \cite{knoll}).

Having computed the diagrams for the process $pn\to pn\: e^+e^-$
we implemented them for the first time in  the reaction $Dp\to p_{\rm sp}\: pn\: e^+e^-$ with
detection of a spectator in the forward direction. The cross section for this
process has been evaluated in a covariant approach based on the Bethe-Salpeter
formalism. Relativistic effects are negligible here.
The Bethe-Salpeter formalism has been used rather for the sake of
consistency with the covariant diagrammatical approach and for convenience in
using the results from  calculations of the subprocess $pn\to pn\: e^+e^-$.
Within the Bethe-Salpeter formalism
we derive a factorization formula, i.e.,
the cross section is cast in a form of a product of two factors, the one entirely
originating from the
deuteron structure and kinematics, the other one being exactly the
cross section of the subprocess $pn\to pn\: e^+e^-$.
In accordance with the factorization formula the shape of the cross section reflects the
one in the elementary reaction, except for some corrections from the deuteron wave function.
Apart from these common features, an essential difference
of the reactions
$Dp\to p_{\rm sp}\: pn\: e^+e^-$ and $np\to \: pn\: e^+e^-$ can appear.
Namely due Fermi motion of nucleons
in the deuteron it is possible to find such kinematical
conditions within those the energy balance is shifted in favor of the
elementary subsystem $np$, so that  subthreshold vector meson production
becomes feasible. However, the performed analysis shows that
the subthreshold cross section is quite low due to a strong suppression
originating from the deuteron wave function.

Finally, we found that the effects of final state interaction can be
neglected at low values of the di-electron mass. With increasing di-electron mass
the final state interaction
effects become more important, in particular at the kinematical limit.

Our results are presented for kinematical conditions accessible
in forthcoming experiments at HADES \cite{HADES,holzman}.
Apart from obtaining  valuable information
for further understanding of di-electron emission in heavy-ion collisions,
this also offers a tool for investigation of
the interplay of bremsstrahlung process and vector meson dominance
effects, i.e., the electromagnetic form factors of nucleons in the time like region
not accessible in experiments with on-mass shell particles
and the properties of the half-off mass shell nucleon-nucleon amplitude.

\section{Acknowledgements}
Fruitful discussions with H.W. Barz, C. Fuchs,
R. Holzmann, J. Knoll and  A.I. Titov are gratefully appreciated.
L.P.K. would like
to thank for the warm hospitality in the Research Center Rossendorf.
This work has been supported by
BMBF grant 06DR121, GSI-FE and the Heisenberg-Landau program.

\appendix

\section{Isospin in the Rarita-Schwinger formalism}\label{isospin}

In full analogy with the spin-$3/2$ space, the wave function
for an isospin $3/2$-particle, e.g., the $\Delta$, can be written as
\be
\chi_{\frac32 \tau_\Delta} =
\sum\limits \la 1\lambda\, \frac12 \tau_N\,|\frac32\tau_\Delta\ra
{\bf e}^*_\lambda\chi_{\frac12\tau_N},
\label{isodelta}
\ee
where $\chi_{\frac32} = col\left(\Delta^{++},\Delta^{+},\Delta^{0},\Delta^{-}\right)$,
$\chi_{\frac12\tau_N}$ is the wave function for isospin-$1/2$, and the
isospin-$1$ wave function ${\bf e}_\lambda\  (\lambda=\pm 1,0)$
has the same structure as in Eq.~($\ref{xilab}$) (see Appendix \ref{a2}).

The definition of the transition matrix $\bf T$ in Eqs.~(\ref{pionDelta})
and (\ref{rhoDelta}) reads
\be
\la \frac32 \tau_\Delta\ |{\bf T}_\alpha|\frac12\tau_N\ra
=\la 1\lambda\, \frac12 \tau_N\,|\frac32\tau_\Delta\ra
\left( {\bf e}^*_\lambda\right )_\alpha .
\label{defineiso}
\ee
Using the completeness of Clebsch-Gordan coefficients one finds
\be
{\bf T}^+_\alpha {\bf T}_\beta = \delta_{\alpha\beta}
-\frac13\boldtau_\alpha \boldtau_\beta .
\label{compl}
\ee
The needed Pauli matrices for isospin 1 are
\be
\begin{array}{ccc}
T_x=\displaystyle\frac{1}{\sqrt{6}}\left(
\begin{array}{lr}
-\sqrt{3}& 0\\
   0     &-1\\
   1     & 0\\
   0 &\sqrt{3}\\
\end{array}\right),\quad
&
T_y=\displaystyle\frac{i}{\sqrt{6}}\left(
\begin{array}{lr}
\sqrt{3}& 0\\
   0     &1\\
   1     & 0\\
   0 &\sqrt{3}\\
\end{array}\right),\quad
&
T_z=\displaystyle\sqrt{\frac{2}{3}}\left(
\begin{array}{lr}
   0& 0\\
   1     &0\\
   0    & 1\\
   0 &0\\
\end{array}
\right) .
\end{array}
\ee
Finally, define charge rising (lowering) operators  $T_{\pm 1}=(T_x\pm iT_y)/2$
with the following properties
\be
\begin{array}{lcl}
T_{+1} \: |p \rangle \sim | \Delta^{++}\rangle , &\phantom{12}&
T_{+1}\: |n\rangle \sim |\Delta^{+}\rangle , \\
T_{-1}\:|p\rangle \sim |\Delta^{0}\rangle , &&
T_{-1}\:|n\rangle \sim |\Delta^{-}\rangle
\end{array}
\ee
to find
\be
\begin{array}{lcrrrcrr}
T_{+1}^+T_z&=&\frac13\tau_{+},& &\phantom{12}& T_{-1}^+T_z&=&-\frac13\tau_{-} ,\\
T_{z}^+T_{+1}&=&-\frac13\tau_{+},& && T_{z}^+T_{-1}&=&\frac13\tau_{-}.
\end{array}
\ee
Note that $\left({\bf T}\boldtau \right) = 2\left( T_{+1}\tau_-
+T_{-1}\tau_+\right ) + T_z\tau_z$ and
\be
\begin{array}{ccc}
T_{z}^+T_z=\displaystyle\frac23
\left(
\begin{array}{cc}
\ 1\ &\ 0\ \\
\ 0\ &\ 1\
\end{array}
\right), &
\tau_+=
\left( \begin{array}{cc}
\ 0\ &\ 1\ \\
\ 0\ &\ 0\
\end{array}
\right) &
\tau_-=\left(\begin{array}{cc}
\ 0\ &\ 0\ \\
\ 1\ &\ 0\
\end{array}
\right)
\end{array}.
\ee
From these equations one finds that the isospin factor for diagonal
(with charge conservation) or non-diagonal (with rising or lowering
of the charge) terms  is $\pm \displaystyle\frac23$. Usually \cite{pascScholten},
in order to simplify the notation, one changes the normalization (\ref{compl}) to
\be
{\bf T}^+_\alpha {\bf T}_\beta = \delta_{\alpha\beta}
-\frac14\left[ \boldtau_\alpha, \boldtau_\beta\right].
\label{compl1}
\ee
In this case the factor $\sqrt{\frac23}$ is included into the coupling
constants. In our calculations we use the normalization (\ref{compl1}).

\section{Two-step model}\label{ts}

In this appendix we establish a correspondence of our diagrammatical
approach to models based on a two-step mechanism according to which
the cross section for di-electron production is expressed as a product of
two terms: (i) production of an on-mass shell
vector meson with the mass around its pole value and  known width, and
(ii) the subsequent decay via the electromagnetic channel
conform known branching ratios.
To this end we recalculate the cross section Eq.~(\ref{diffcross}) within
the VMD conjecture and cast it
in a form close to a two-step model. For definiteness, let us calculate
the diagram a) in Fig.~\ref{fig2}, where now the photon couples
the nucleon via an isoscalar vector meson, e.g., the $\omega$ meson. 
The corresponding part of the amplitude $T$ reads
\begin{eqnarray}
T(q^2)&=&
\bar u({\bf p}_1',s_1') \hat O_{mes}(k,P_2,s_2,P_2',s_2') S(P_1-q)\Gamma^{\mu(\omega)}
u({\bf p}_1,s_1)\nonumber\\&   \times&
\displaystyle\frac{-g_{\mu\mu'}+\displaystyle\frac{q_\mu q_{\mu'}}{q^2}}{q^2-M_0^2}
\left(\frac{ M_{0}^2}{f_{\omega\gamma}}\right)
\frac{e^2 g_{\mu'\mu''}}{q^2} \bar u(k_1,\tilde s_1)\gamma^{\mu ''} v(k_2,\tilde s_2),
\label{b1}
\end{eqnarray}
where formally the operator $\hat O_{mes}(k,P_2,s_2,P_2',s_2')$ includes all
the exchange mesons, meson propagators and contributions from the lower
vertex Fig.~\ref{fig2}a. For further convenience,
in the vector  propagator in Eq.~(\ref{b1}) we replace the quantity
$\displaystyle\frac{q_\mu q_\mu'}{M_0^2}$
by $\displaystyle\frac{q_\mu q_\mu'}{q^2}$ (since due to  gauge invariance these
terms do not contribute to the amplitude (\ref{b1})).
Note that Eq.~(\ref{b1}) is
valid at any value of $q^2$, which generally, could be quite far from the
pole mass $M_0^2$. Let us introduce a hypothetical on-mass shell
particle with the invariant mass $s_V=s_\gamma=q^2$ and with quantum numbers
as those of $\omega$.
Then, for such a particle one can use the completeness relation for
its polarization vectors $\xi_\lambda (q^2) $ ($\lambda = 0,\pm 1$) to write
\be
\sum\limits_\lambda
\:\xi_\lambda^\mu (q^2) \:\xi^{+ \mu'}_\lambda (q^2) =
-g^{\mu\mu'}+\displaystyle\frac{q^\mu q^{\mu'}}{q^2}
\label{b2}
\ee
and
\begin{eqnarray}
\!\!\!\!\!\!\!\!\!\!\!\!
T(q^2)& =&
\sum\limits_\lambda \left[
\bar u({\bf p}_1',s_1') \hat O_{mes}(k,P_2,s_2,P_2',s_2') S(P_1-q)
\left( \Gamma^{(\omega)}\cdot \xi_{\lambda}^+\right)
u({\bf p}_1,s_1) \right]\nonumber\\&\times&
\displaystyle\frac{1}{q^2-M_0^2}
\left( \frac{ M_{0}^2}{f_{\omega\gamma}} \right)
\frac{e^2}{q^2}  \bar u(k_1,s_1)\left(\xi_\lambda\cdot\gamma\right) v(k_2,s_2)
\nonumber\\& \equiv&
\sum\limits_\lambda {\cal A}_{NN\to NNV} \left(q^2 \neq M_0^2, \lambda\right)
\displaystyle\frac{1}{q^2-M_0^2}
\left( \frac{ M_{0}^2}{f_{\omega\gamma}} \right)
\frac{e^2}{q^2}  \bar u(k_1,s_1)\left(\xi_\lambda\cdot\gamma\right) v(k_2,s_2),
\label{b3}
\end{eqnarray}
where ${\cal A}_{NN\to NNV} \left(q^2 \neq M_0^2, \lambda\right)$ corresponds
to the production amplitude of a vector particle with invariant mass
$q^2$ and polarization $\lambda$.
The obtained formula has almost the desired factorized form, however,
due  the dependence upon $\lambda$ of both terms in (\ref{b3}), the factorization is
not yet complete. The exact factorization can be accomplished for the squared
amplitude after summation over spins and integrating over the leptonic phase space,
\begin{eqnarray}
&&
\frac14\sum_{spins}\: \int \left | T(q^2)\:\right|^2 R_2(q\to k_1+k_2) =
\nonumber\\&&
\frac14\sum_{spins}\sum_{\lambda\lambda'}
\bar u({\bf p}_1',s_1') \hat O_{mes}(k,P_2,s_2,P_2',s_2') S(P_1-q)
\left( \Gamma^{(\omega)}\cdot \xi_{\lambda}\right) u({\bf p}_1,s_1)\:
\nonumber\\&&\times
\bar u({\bf p}_1,s_1)\left( \Gamma^{(\omega)}\cdot \xi_{\lambda'}^+\right)
 S(P_1-q) \hat O_{mes}^+(k,P_2,s_2,P_2',s_2')
u({\bf p}_1',s_1')
\nonumber\\&&\times
\left| \frac{1}{\left( q^2-M_0^2\right)}\right|^2
\frac{e^4}{q^4}\int R_2(q\to k_1+k_2)
\left( \frac{ M_{0}^2}{f_{\omega\gamma}} \right)^2
\xi_\lambda^\mu\:
l_{\mu\nu}(k_1,k_2)\:\xi_{\lambda'}^{+\nu}.
\label{b4}
\end{eqnarray}
Carrying out the integration over $R_2(q\to k_1+k_2)$ and observing that it provides
a $\delta_{\lambda\lambda'}$ function one gets
\begin{eqnarray}&&
\frac14\sum_{spins}\: \int  \left | T(q^2)\:\right|^2 R_2(q\to k_1+k_2)=
\nonumber\\&&\
\frac14\sum_{spins}\sum_{\lambda}
\left| {\cal A}_{NN\to NNV} \left(q^2 \neq M_0^2, \lambda\right)\right |^2
\frac{8\pi^2 \sqrt{s_\gamma}\:\Gamma_{em}(q^2)}{(q^2-M_0^2)^2  + M_0^2 \Gamma_{tot}^2(q^2)},
\label{b5}
\end{eqnarray}
where the quantity
\be
\Gamma_{em}(q^2)=\frac{4\pi \alpha_{em}^2}{3s_\gamma\sqrt{s_\gamma}}
\left( \frac{ M_{0}^2}{f_{\omega\gamma}} \right)^2
\label{b6}
\ee
plays the role of the electromagnetic decay width of a vector particle
into a di-electron  via intermediate creation of a virtual photon.
Then the cross section %(\ref{sechenie}) 
can be written in the form
\begin{eqnarray}
\frac{d\sigma}{ds_\gamma} &=&
\left [\frac{1}{2(2\pi)^5\sqrt{\lambda(s,m^2,m^2)}}
\frac14\sum_{spins}\int d s_{12} R_2(P_1+P_2\to P_V+P_{12})
R_2(P_{12}\to P_1'+P_2')\right.\nonumber\\&\times&\left.
\left| {\cal A}_{NN\to NNV} \left(q^2 \neq M_0^2,
\lambda\right)\phantom{\frac{1^2}{2^2}}\right |^2\right]
\frac{\sqrt{s_\gamma}\: \Gamma_{em}(q^2)/\pi}{(q^2-M_0^2)^2  + M_0^2 \Gamma_{tot}^2(q^2)},
\label{b7}
\end{eqnarray}
\noindent
where the expression in square brackets can be interpreted as
cross section for the creation
of a vector particle with the mass $s_\gamma=q^2$ in a $NN$ process,
\begin{eqnarray}
\frac{d\sigma}{ds_\gamma}=
 \sigma_V(q^2\neq M_0^2)\:
\frac{ \sqrt{s_\gamma}\:\Gamma_{em}(q^2)/\pi}{(q^2-M_0^2)^2  + M_0^2 \Gamma_{tot}^2(q^2)}.
\label{b8}
\end{eqnarray}
It is worth emphasizing that
the invariant mass $q^2$ varies in the whole kinematical
range,  $0<\:q^2\: <\: q^2_{max}$. Hence, the form of the cross section (\ref{b8})
is valid at any initial energy, including deep-subthreshold values, $q^2_{max}\ll M_0^2$.
Therefore, at such invariant masses the cross section can be considered
as background contribution to di-electron production.
In the very vicinity of the pole masses, $q^2\to M_0^2$, the cross section (\ref{b8})
becomes
\begin{eqnarray}
\frac{d\sigma}{ds_\gamma}=
\sigma_\omega (M_0^2)\:
\frac{ \Gamma_{em}/\left(4\pi \sqrt{s_\gamma}\right)}
{\left(\sqrt{s_\gamma}-M_0\right)^2  +  \Gamma_{tot}^2/4}
\label{b9}
\end{eqnarray}
which exactly coincides with results of the two-step model.
Remind  that our diagrammatical approach can be related to the
two-step model only if (i) one considers the cross section
integrated over the di-electron phase space, (ii) the invariant mass
$s_\gamma$ is not too far from the pole masses, (iii) interferences
between different kinds of vector mesons, $\omega$ and $ \rho$,
are disregarded, and (iv) the contributions of other diagrams (e.g., with
$\Delta$ isobars) are  neglected.

\section{Factorization}\label{a2}

The BS amplitudes in the deuteron rest system are of the
form \cite{quad,Tjon}
\begin{eqnarray}
\Psi_{\CM_D}^{S^{++}}(P_n,p_{\rm sp})&=&
{\cal N}(\hat k_n+m )\frac{1+\gamma_0}{2}\hat\xi_{\CM_M}(\hat p_{\rm sp}-m)
\phi_S (p_0,|{\bf P}_n|), \label{psis}\\[2mm]
\Psi_{\CM_D}^{D^{++}}(P_n,p_{\rm sp})&=&-\frac{{\cal N}}{\sqrt{2}}
(\hat k_n+m )\frac{1+\gamma_0}{2}
\left(
\hat\xi_{\CM_D} +\frac{3}{2|{\bf P}_n|^2} (\hat k_n-\hat p_{\rm sp})(p\xi_M)\right) \nonumber\\
&& \times (\hat p_{\rm sp}-m) \phi_D (p_0,|{\bf P}_n|),\nonumber
%\label{psid},
\end{eqnarray}
where
$k_n$ is an on-mass shell four vector related  to the
off-mass shell neutron vector $P_{n}$ as follows
(in anti-laboratory system one has ${\bf P}_n=-{\bf p}_{\rm sp}$)
\begin{equation}
k_n=(E_{{\bf k}_n},{\bf P}_n),\quad E_{{\bf k}_n}=\sqrt{{\bf P}_n^2+m^2};\quad
p=(p_0,{\bf P}_n),\quad p_0=\frac12 M_D - E_{\rm sp}.
\end{equation}
$\phi_{S,D} (p_0,|{\bf P}_n|)$ are the partial scalar amplitudes related to
the corresponding partial vertices as
\begin{equation}
\phi_{S,D} (p_0,|{\bf P}_n|)=
\displaystyle\frac{G_{S,D}
(p_0,|{\bf P}_n|)}{\left(\displaystyle\frac{M_D}{2}-E_{\rm sp} \right)^2-p_0^2}.
\label{verBS}
\end{equation}
In Eq.~(\ref{psis}) $M_D$ is the deuteron mass,and the normalization factor is
${\cal N}=\displaystyle\frac{1}{\sqrt{8\pi}}
\displaystyle\frac{1}{2E_{\rm sp}(E_{\rm sp}+m)}$.

The components of the
polarization  vector of a vector particle moving with
4-momentum  $p=(E,{\bf p})$ having the  polarization
projection $\CM=\pm 1,\, 0$ and  mass $M$ are
\begin{eqnarray}
&&
\xi_\CM=\left( \frac {\bf p \boldxi_\CM}{M},
\boldxi_\CM+{\bf p}
\frac {\bf p \boldxi_\CM}{M(E_{\bf p}+M)} \right ),
\label{xi}
\end{eqnarray}
where $\boldxi_\CM$ is the polarization vector for the particle at rest with
\begin{equation}
\begin{array}{ccc}
\boldxi_{+ 1} =-\displaystyle\frac{1}{\sqrt{2}}\,
\left ( \begin{array}{c}1\\ i\\0 \end{array} \right ),
&
\quad \boldxi_{- 1}=\displaystyle\frac{1}{\sqrt{2}}
\left ( \begin{array}{c}1\\ -i\\0 \end{array} \right ),
&
\quad \boldxi_{0}=
\left ( \begin{array}{c} 0\\ 0\\1 \end{array} \right ).
\label{xilab}
\end{array}
\end{equation}
The Dirac spinors, normalized  as $\bar{u}(p) u(p)=2m$ and $\bar{v}(p) v(p)=-2m$, read
\begin{eqnarray}
u({\bf p},s)=\sqrt{m+ E_{{\bf p}}}
\left ( \begin{array}{c} \chi_s \\
\displaystyle\frac{\boldsigma {\bf p}}{m+ E_{{\bf p}}}\chi_s
\end{array}  \right), \quad \quad
v({\bf p}\,,s)=\sqrt{m+ E_{{\bf p}}}\left( \begin{array}{c}
\displaystyle\frac{\boldsigma {\bf p}}{m+ E_{{\bf p}}}\widetilde\chi_{s}\\
\widetilde\chi_{s}
\end{array}  \right),
\label{spinors}
\end{eqnarray}
where $\widetilde\chi_s\equiv -i\sigma_y\chi_s$,  and
$\chi_s$ denotes the usual two-dimensional Pauli spinor.
Note that the denominator in (\ref{verBS}) is zero when one particle
(the spectator in our case) is on-mass shell. This singularity is
apparent since it is exactly compensated by the factors
$(\hat p_{\rm sp}-m)\ \tilde S^{-1}(p_{\rm sp}) = p_{\rm sp}^2-m^2 = 0$ from
(\ref{psis}) and (\ref{deucurr}).
In terms of the main BS components
the ''deuteron spinor'' ${\cal V}_D$ (\ref{vd}) can be written in as
\be
{\cal V}_D= {\cal N} (p_{\rm sp}^2-m^2) \ \left( \phi_S V_S
 -\frac{1}{\sqrt{2}}\phi_D V_S-\frac{1}{\sqrt{2}}\phi_D\ V_D\right),
\label{spinor1}
\ee
where
\begin{eqnarray}
V_S&=&(\hat k_n+m) \left(
\begin{array}{cc}
1\ & 0\ \\ 0\ & 0\
\end{array}\right)
\left( \begin{array}{cc}
0\ & -(\boldsigma\boldxi)\ \\ (\boldsigma\boldxi)\ & 0\
\end{array}\right)
\sqrt{E_{\rm sp}+m}
\left(\begin{array}{c}
-\displaystyle\frac{(\boldsigma {\bf P}_n)}{E_{\rm sp}+m}\widetilde\chi_{s} \\
\widetilde\chi_{s}
\end{array}\right)\label{spvs} \nonumber\\
 &=& \sqrt{E_{\rm sp}+m}\left(  \begin{array}{cc}
 0\ & -(E_{\rm sp}+m)(\boldsigma\boldxi)\ \\
 0\ & -({\bf P}_n\boldsigma)(\boldsigma\boldxi)
 \end{array}\right)
 \left(\begin{array}{c}
 -\displaystyle\frac{(\boldsigma {\bf P}_n)}{E_{\rm sp}+m}\widetilde\chi_{s}\\[2mm]
 \widetilde\chi_{s}
 \end{array}\right) \nonumber\\
&=&   -\left({E_{\rm sp}+m}\right)^{3/2}
\left(  \begin{array}{c}
(\boldsigma\boldxi)\widetilde\chi_{s}\ \\[2mm]
\displaystyle\frac{({\bf P}_n\boldsigma)(\boldsigma\boldxi)}
{E_{\rm sp}+m}\widetilde\chi_{s}
\end{array}\right) ,
\end{eqnarray}
\begin{eqnarray}
V_D&=&\displaystyle\frac{3({\bf P}_n\boldxi)}{{\bf P}_n^2}
\left( \begin{array}{cc} E_{\rm sp}+m\ & -({\bf P}_n\boldsigma)\\
({\bf P}_n\boldsigma)\ &-E_{\rm sp}+m
\end{array}\right)
\left(\begin{array}{cc} 1\ &0 \\0\ &0\end{array}\right)
\left( \begin{array}{cc} 0 & ({\bf P}_n\boldsigma)\\
- ({\bf P}_n\boldsigma)\ &0
\end{array}\right) \nonumber\\[2mm]
&&\times \sqrt{E_{\rm sp}+m} \left(\begin{array}{c}
-\displaystyle\frac{(\boldsigma {\bf P}_n)}{E_{\rm sp}+m}\widetilde\chi_{s}\\
\widetilde\chi_{s}
\end{array}\right) \nonumber\\
&=& \left(E_{\rm sp}+m\right)^{3/2}\displaystyle\frac{3({\bf P}_n\boldxi)}{{\bf P}_n^2}
\left(  \begin{array}{c}
({\bf P}_n\boldsigma)\widetilde\chi_{s} \\[2mm]
\displaystyle\frac{{\bf P}_n^2}{E_{\rm sp}+m} \widetilde\chi_{s}
\end{array} \right) .
\label{vspe}
\end{eqnarray}
From (\ref{spvs}) and (\ref{vspe}) one finds
\be
\sum\limits_{\CM_D,s} V_S\bar V_S = 3(E_{\rm sp}+m)^2(\hat k_n+m)\label{ss},\\&&
\sum\limits_{\CM_D,s} V_D\bar V_D = 9(E_{\rm sp}+m)^2(\hat k_n+m)\label{dd},\\&&
\sum\limits_{\CM_D,s} V_S\bar V_D=\sum\limits_{\CM_D,s} V_D\bar V_S
=-3(E_{\rm sp}+m)^2(\hat k_n+m). \label{ds}
\ee
Now it is straightforward to obtain (\ref{vv}) from
Eqs.~(\ref{psis}) - (\ref{ds}). In Eq.~(\ref{vv}) the deuteron wave functions
$U_{S,D}(|{\bf p}_{\rm sp}|)$ are related with the half off-mass shell vertices
$G_{S,D} $ (\ref{verBS}) as
\be
U_{S,D}(|{\bf p}_{\rm sp}|)=\displaystyle\frac
{G_{S,D}(p_0 = \frac12 M_D-E_{\rm sp}, |{\bf p}_{\rm sp}|)}
{4\pi\sqrt{2M_D}\left( M_D-2E_{\rm sp}\right)}.
\ee

\newpage

%%%%%%%%%%%%%%%%%%%%%%%%%%%%%%%%%%%%%%%%%% BEGIN FIGURES  %%%%%%%%%%%%%%%%%

\begin{table}[h]

\begin{tabular}{c c c c ccc }
  \hline\hline  % after \\: \hline or \cline{col1-col2} \cline{col3-col4} ...
 Meson          &\phantom{ppp} &   $\displaystyle\frac{g^2_{NNM}}{4\pi}$&\phantom{qqq}$\kappa$\phantom{qqq} &     l       & &  $\Lambda$      \\
                &              &                           &    & ($GeV^{-1}$)& &   (GeV)        \\\hline
    $\pi $      &              &12.562             & -     & 0.1133           & &  1.005         \\
    $\sigma$    &              &  2.340            & -     & 0.1070           & & 1.952 \\
    $\rho$      &              & 0.317             & 6.03  & 0.1800           & &  1.607    \\
    $\omega$    &              &46.035             &  0    & 0.0985           & & 0.984     \\
  \hline\hline
\end{tabular}
\caption{OBE parameters used in the calculations 
(cf.\ Eq.~(\ref{obeMosel}) and Ref.~\cite{mosel_calc}). Note that
$f_{  NN\pi }=\displaystyle\frac{\mu_\pi }{2m}g_{ NN\pi} $.}
\label{tabl1}
\end{table}

\begin{figure}[h]  %      Fig1  chu-low diagrams
\includegraphics[width=0.999\textwidth]{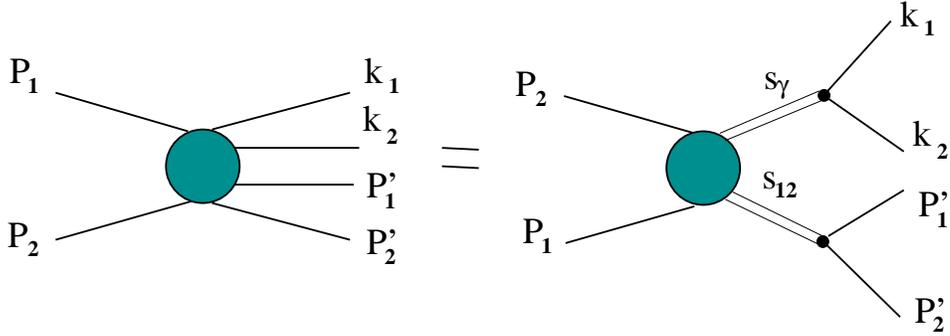} %
\caption{Choice of kinematical variables for the process
$N_1(P_1)+ N_2(P_2) \to N_1'(P_1') + N_2'(P_2') + e^+(k_1) + e^-(k_2)$.}
\label{fig1}
\end{figure}

\begin{figure}[h]  %      Fig2 Feynman diagrams for NN  bremsstrahlung
\includegraphics[width=1.1\textwidth]{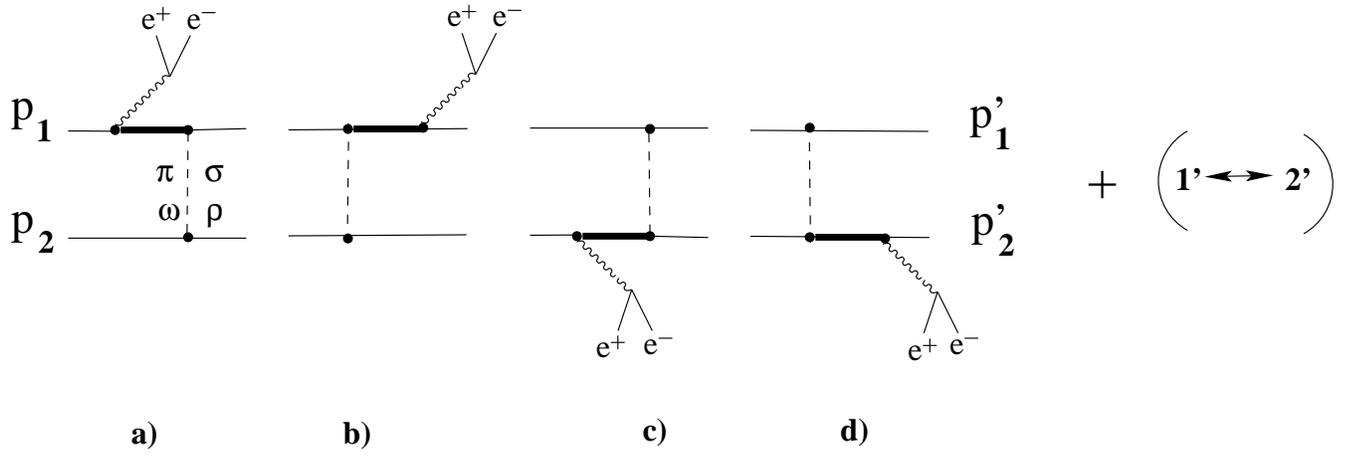} %
\caption{Bremsstrahlung diagrams for the process
$N_1+N_2 \to N_1' + N_2' + e^+ e^-$ in the one-photon and one-boson exchange
approximation (thick lines denote propagators of either
nucleons or baryon resonances).}
\label{fig2}
\end{figure}

\begin{figure}[h]  %      Fig3      MEC
~\vskip3cm
\includegraphics[width=1.0\textwidth]{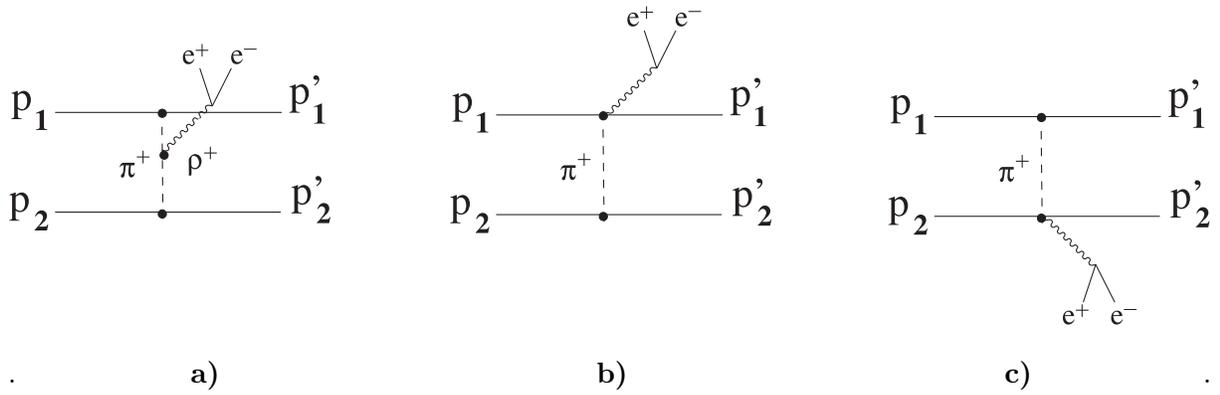} %\hspace*{5mm}
. {\bf \hspace*{2cm} a) \hfill b) \hfill c) \hspace*{2cm}} .\\
\caption{Contribution of meson exchange currents (a)
and seagull terms (b, c)
to the  process  $N_1+N_2 \to N_1' + N_2' + e^+ e^-$,
where $N_1$ and  $N_2'$ stand for protons and
$N_2$ and $N_1'$ denote neutrons.}
\label{fig3}
\end{figure}

\begin{figure}[h]  %      Fig4  pp at "Mosel" energies
\includegraphics[width=0.999\textwidth]{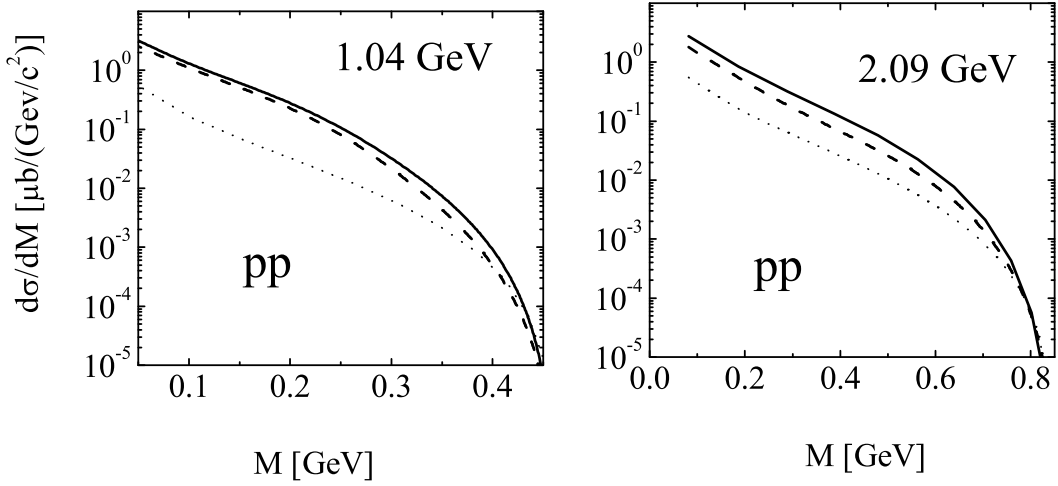} %
\caption{Invariant mass distribution of a $e^+ e^-$ pair in
proton-proton collisions. The dashed (dotted)
curves depict the contribution of diagrams with
bremsstrahlung from $\gamma \Delta N$ ($\gamma NN$)
vertices. The solid lines are the results of calculations
of the total cross section as coherent sums of nucleon and $\Delta$
contributions. The calculations have been performed at DLS energies
\cite{DLS} and are directly comparably with results of \cite{mosel_calc}.}
\label{fig4}
\end{figure}

\begin{figure}[h]  %      Fig5 pn at "Mosel" energies
\includegraphics[width=0.999\textwidth]{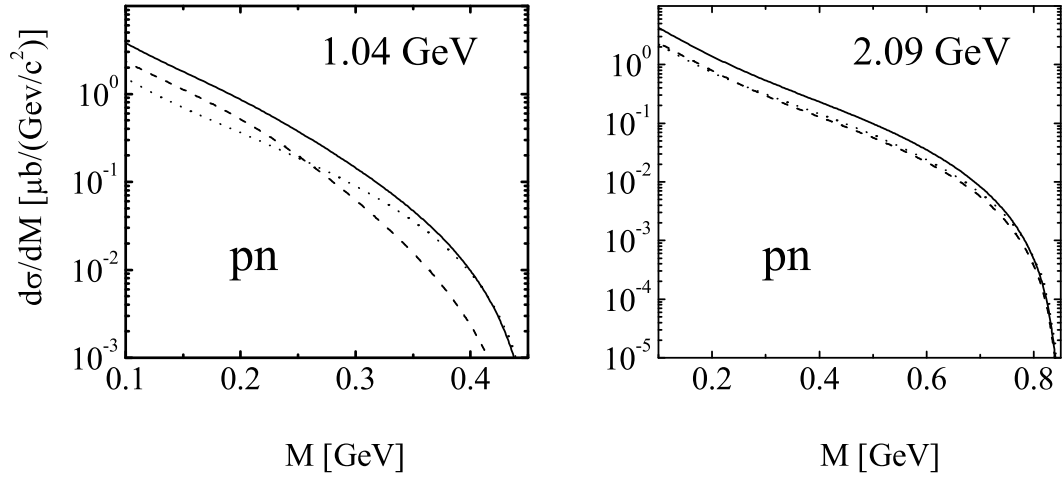} %
\caption{The same as in Fig.~\ref{fig4} but for $pn$ reactions. 
The dotted curves include also contributions from $\gamma \pi N N$ vertices.}
\label{fig5}
\end{figure}

\begin{figure}[h]  %      Fig6 pp at "HADES" energies
\includegraphics[width=0.999\textwidth]{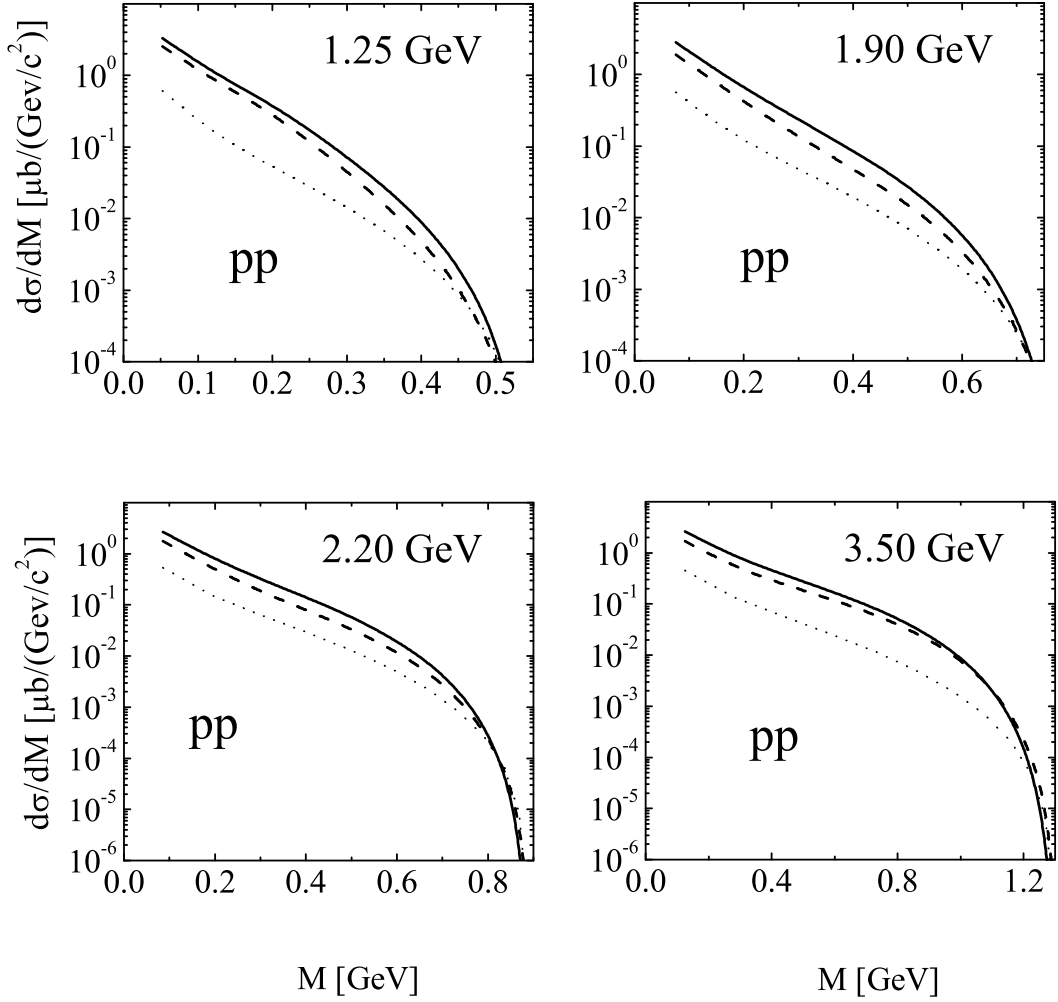} %
\caption{Invariant mass distribution of di-electrons in $pp$ reactions
at energies envisaged in experiments at HADES \cite{HADES}.
Notation as in Fig.~\ref{fig4}.}
\label{fig6}
\end{figure}

\begin{figure}[h]  %      Fig7 pn at "HADES" energies
\includegraphics[width=0.999\textwidth]{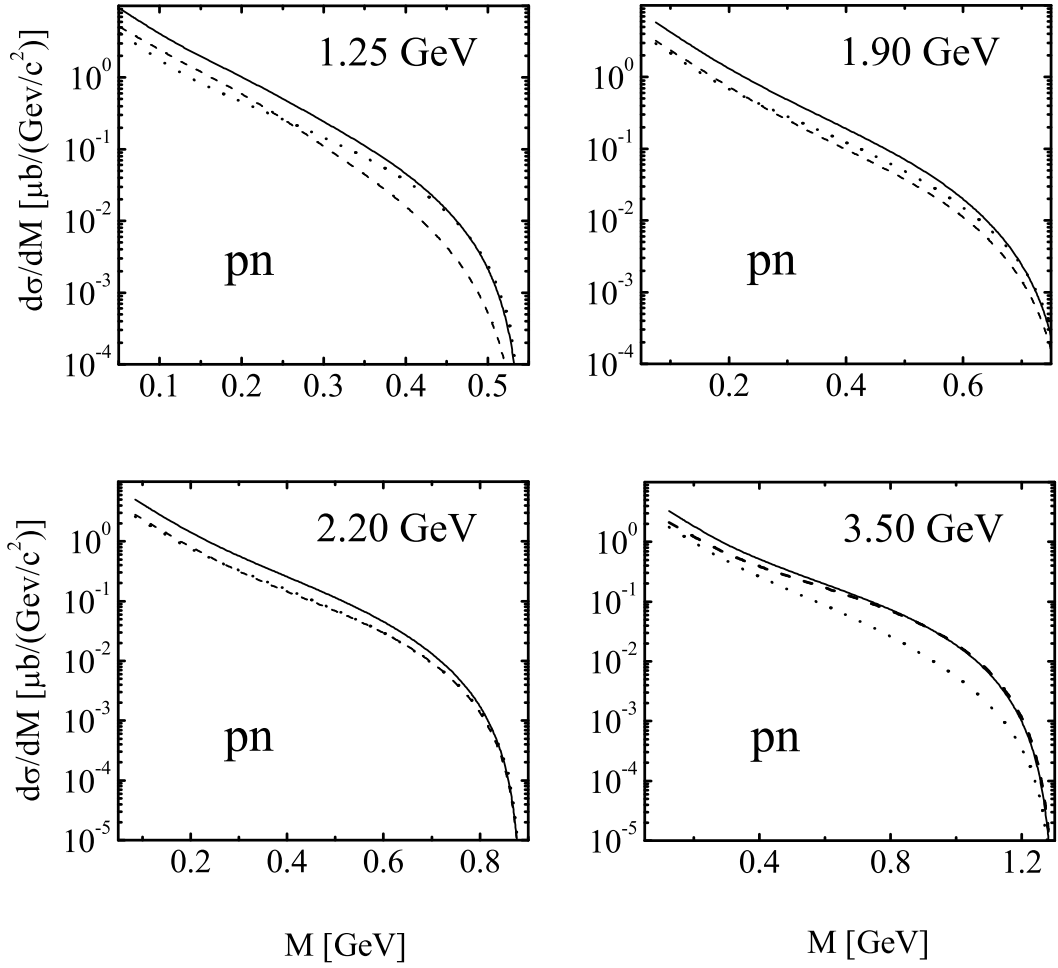} %
\caption{The same as in Fig.~\ref{fig6} but for $pn$ reactions. }
\label{fig7}
\end{figure}

\begin{figure}[h]  %      Fig8 Iso Ratio  at "HADES" energies
\includegraphics[width=0.999\textwidth]{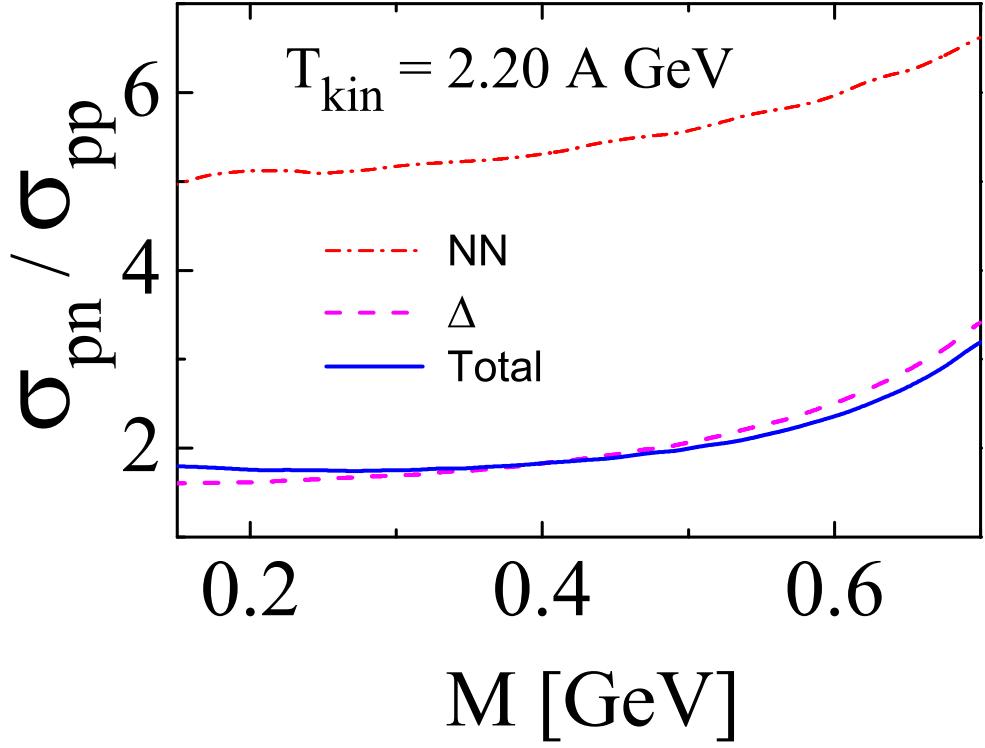} %
\caption{
The ratio $\sigma ( pn \to pn e^+ e^-)$ to $\sigma ( pp \to pp e^+ e^-)$
as a function of the invariant mass at $T_{\rm kin} = 2.2 \, GeV$.
The dot-dashed and dashed curves exhibit results
with only nucleon and $\Delta$ contributions, respectively. The ratio
of the total cross sections is denoted by the solid line.}
\label{fig8}
\end{figure}

\begin{figure}[h]  %      Fig9           %VMD in pp
\includegraphics[width=0.999\textwidth]{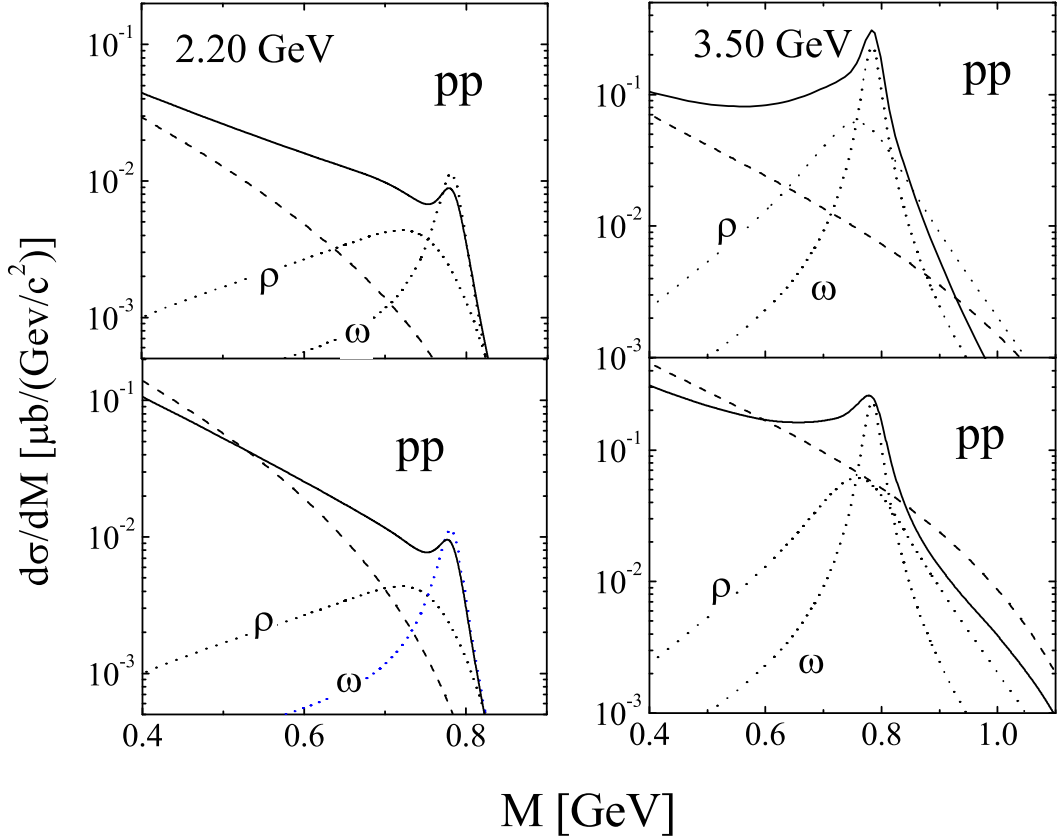} %
\caption{Illustration of the role of the VMD effects for the
invariant mass distribution of di-electrons produced in $pp$ reactions.
The left and right panels are for
initial energies $T_{kin}=2.2\: GeV$ and $T_{kin}=3.5\: GeV$, respectively.
In the upper row results are presented for
nucleon contributions solely, while in the lower one the
total cross sections, including $\Delta$ isobars, are shown.
The dashed lines present the background contribution (cf.\ Fig.~\ref{fig6}),
dotted lines exhibit the pure VMD effects, i.e., the separate contribution
from $\rho$ and $\omega$ mesons. The solid lines illustrate the effects
of VMD for the total cross section (bottom panels) and
for pure nucleon contributions (i.e., without $\Delta$, top panels).}
\label{fig9}
\end{figure}

\begin{figure}[h]  %      Fig10           %VMD in pn
\includegraphics[width=0.999\textwidth]{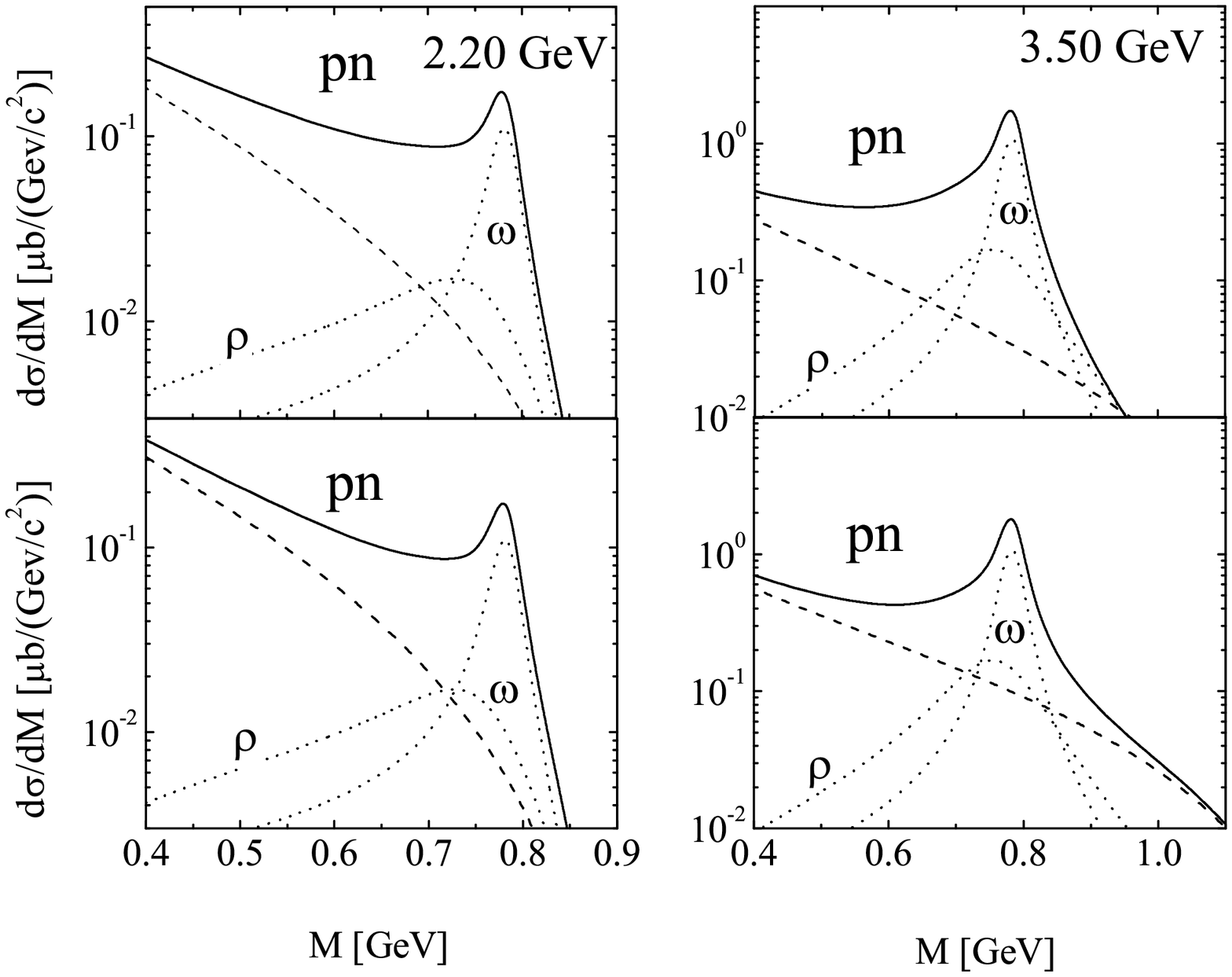} %
\caption{Same as Fig.~\ref{fig9} but for $pn$ reactions.}
\label{fig10}
\end{figure}

\begin{figure}[h]  %      Fig10a           %VMD+FSI in pn
\includegraphics[width=0.999\textwidth]{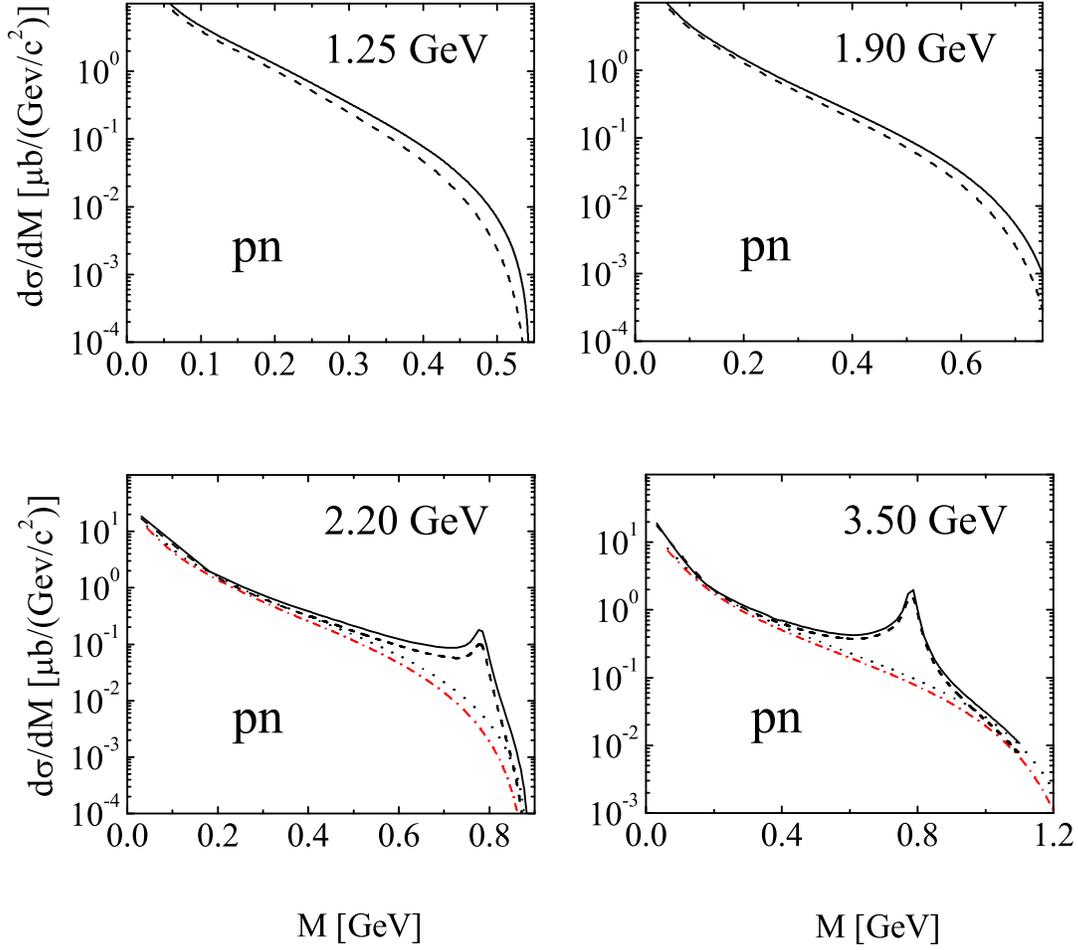} %
\caption{ Illustration of the role of
FSI effects in the invariant mass distribution of di-electrons
produced in $pn$ reactions.
The four panels correspond to the
initial beam energies $T_{kin}=1.25\ GeV$,
$\: 1.9\ GeV$, $\: 2.2\ GeV$ and $\: 3.5\ GeV$, respectively.
The dashed lines present the contribution without FSI,
while the solid lines illustrate the effects
of FSI. In the lower row effects
of VMD are displayed as well: dot-dashed lines are results of
background contribution without FSI,
dotted lines exhibit the  background including FSI.}
\label{fig10a}
\end{figure}

\begin{figure}[h]  %      Fig11              diagramma Dp->eeNN
\includegraphics[width=0.999\textwidth]{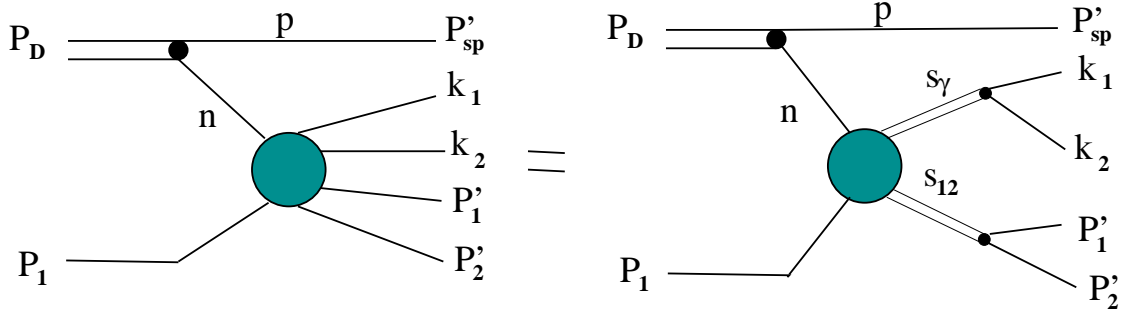} %
\caption{ Kinematics of  the process
$D(P_D) + p(P_1) \to p(P_{\rm sp} + p(P_1') + n(P_2') + e^+(k_1) + e^-(k_2)$
within the spectator mechanism.}
\label{fig11}
\end{figure}

\begin{figure}[h]  %      Fig12   momentum distribution  vs. spectator momentum at fixed M
\includegraphics[width=0.999\textwidth]{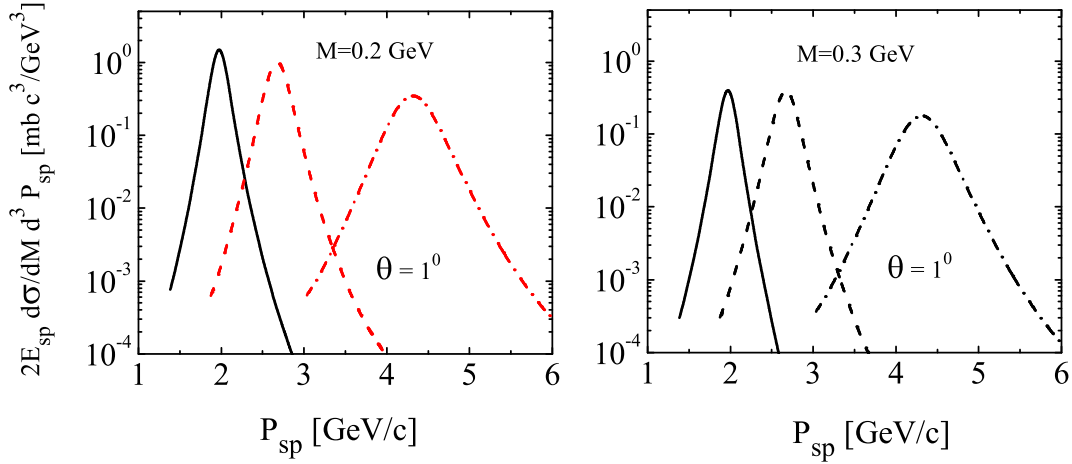} %
\caption{ Dependence of the differential cross section
$2E_{\rm sp}\displaystyle\frac{d\sigma}{dM d^3p_{\rm sp}}$ for
the reaction $Dp\to p_{\rm sp}\ np\ e^+e^-$
on the spectator momentum at two values
of the invariant mass of the lepton pair, $M=0.2 \ GeV$ (left panel)
and $M=0.3 \ GeV $ (right panel). The solid,
dashed and dot-dashed lines correspond to beam energies
$T_{\rm kin}=1.25\ A GeV$, $T_{\rm kin}=1.90\ A GeV$
and $T_{\rm kin}=3.50 \ A  GeV$
respectively. The spectator is assumed to be detected at $\theta = 1^\circ$ 
in the laboratory frame.}
\label{fig12}
\end{figure}

\begin{figure}[h]  %      Fig13  mass distribution at fixed Psp vs invariant mass
\includegraphics[width=0.999\textwidth]{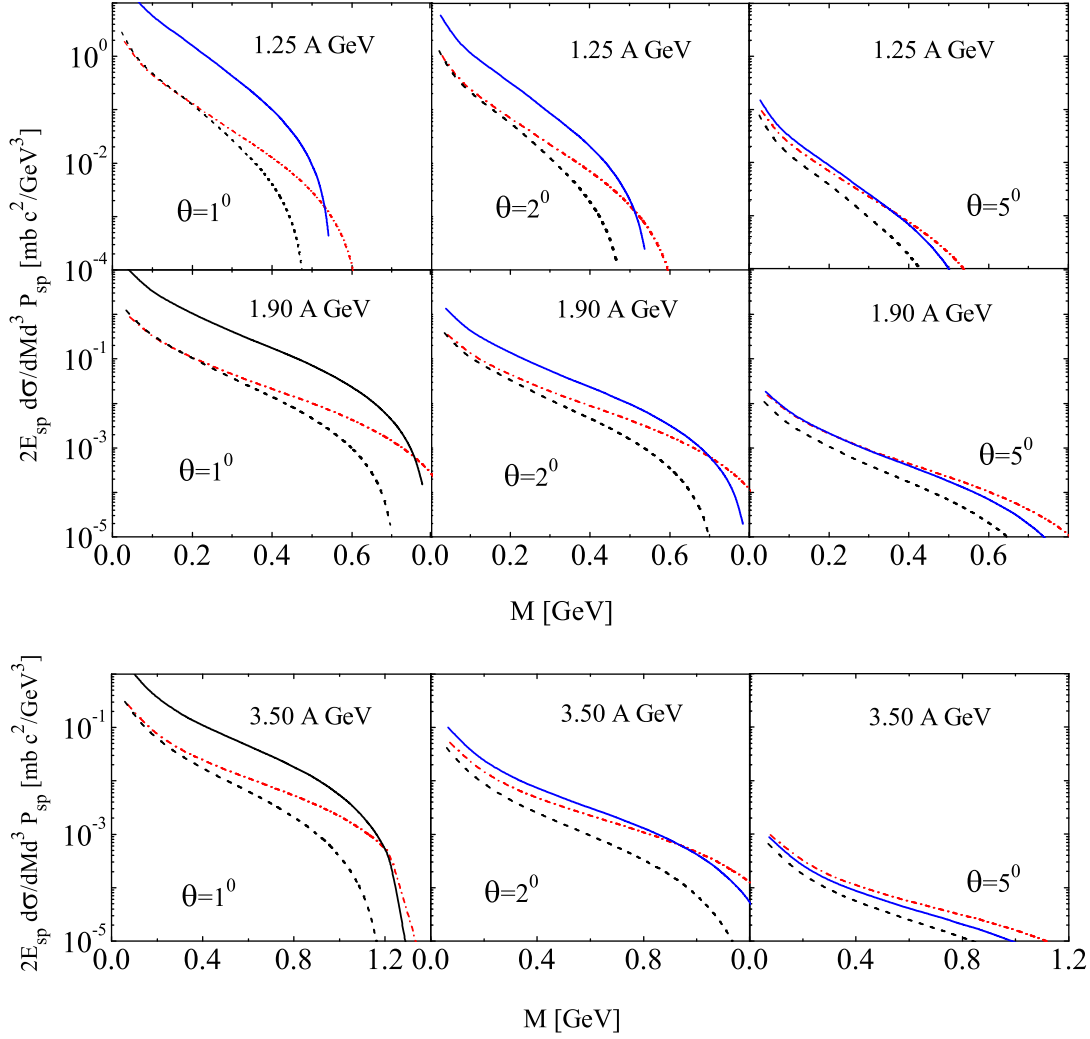} %
\caption{Invariant mass
distribution $2E_{\rm sp}\displaystyle\frac{d\sigma}{dM d^3p_{\rm sp}}$
for the reaction $Dp\to p_{\rm sp}\ np\ e^+e^-$
for three values of the deuteron beam energy,
$T_{kin}=1.25\  A\: GeV$ (upper panel), $T_{kin}=1.90\ A GeV$ (middle panel)
and $T_{kin}=3.5 \ A\: GeV$ (lower panel) and three
values of the spectator angle in the laboratory system,
$\theta= 1^\circ$ (left column), $\theta= 2^\circ $(middle column)
and $\theta= 5^\circ$ (right column).
Dot-dashed, solid and dashed curves correspond
to values of the spectator momentum $\bsp=0.45 \bpD$,
$\bsp=0.50 \bpD$ and $\bsp=0.55 \bpD$, respectively.}
\label{fig13}
\end{figure}

\begin{figure}[h]  %      Fig14   VMD in the deuteron
\includegraphics[width=0.999\textwidth]{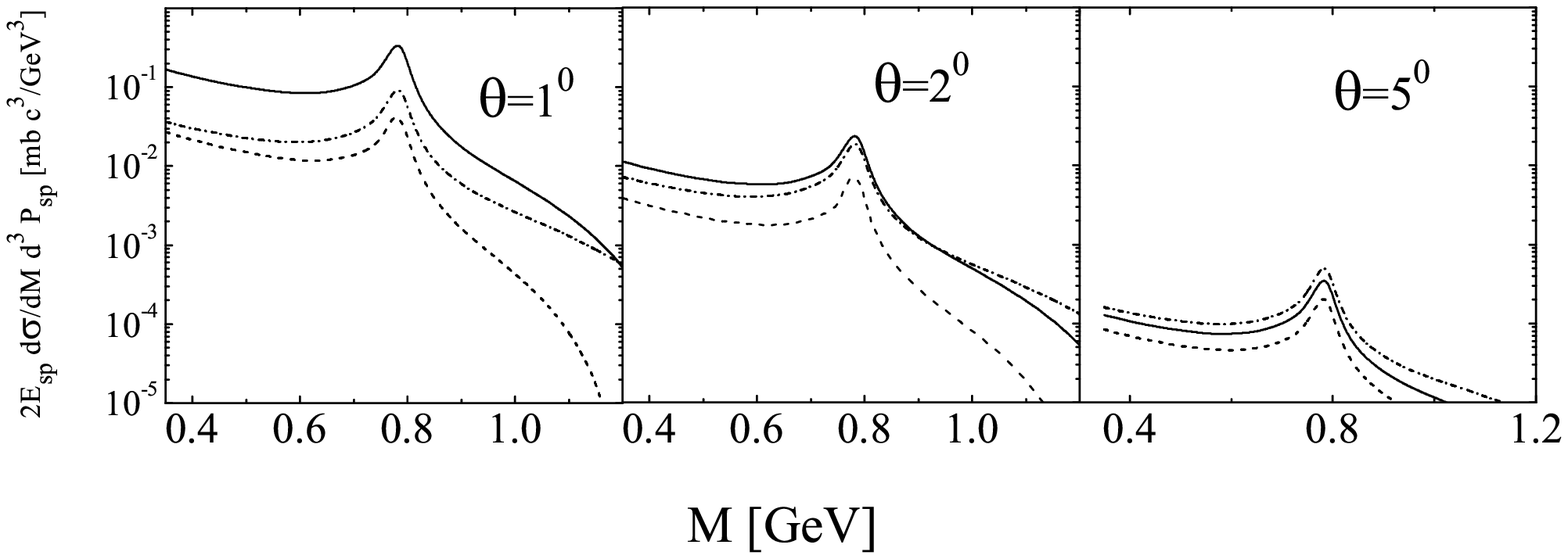} %
\caption{Invariant mass
distribution $2E_{\rm sp}\displaystyle\frac{d\sigma}{dM d^3p_{\rm sp}}$
for the reaction $Dp\to p_{\rm sp}\ np\ e^+e^-$,
at deuteron beam energy $T_{kin}=3.5 \ A GeV$ and three
values of the spectator angle in the laboratory system,
$\theta= 1^\circ, 2^\circ$ and $5^\circ$.
Dot-dashed, solid and dashed curves correspond
to values of the spectator momentum $\bsp=0.45 \bpD$,
$\bsp=0.50 \bpD$ and $\bsp=0.55 \bpD$, respectively.
Effects of VMD have been taken into account.}
\label{fig14}
\end{figure}

\begin{figure}[h]  %      Fig16   SUb-threshold VMD
\includegraphics[width=0.999\textwidth]{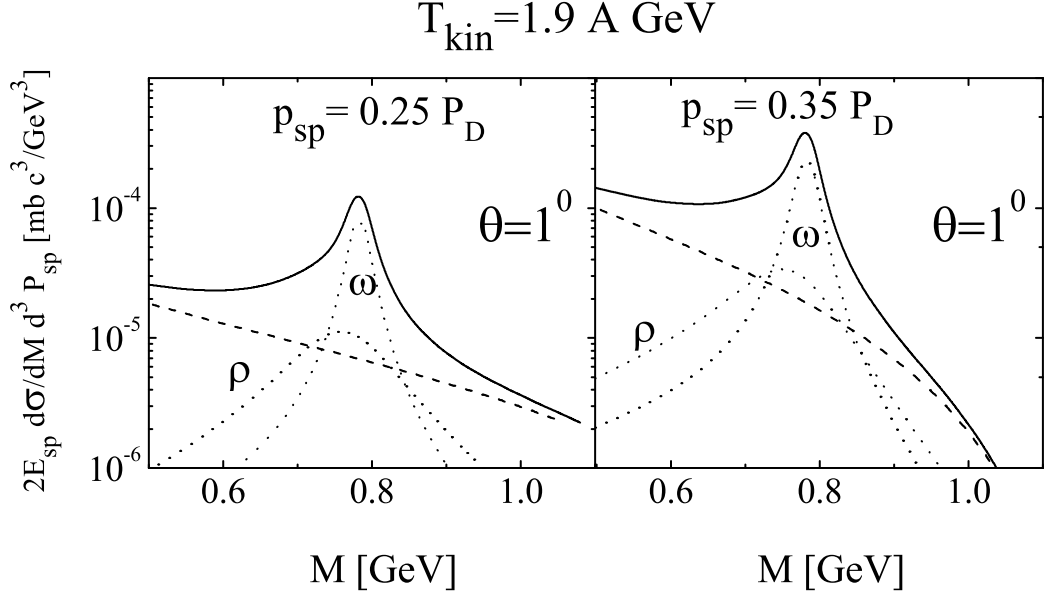} %
\caption{Effects of sub-threshold vector meson
production in the reaction $Dp\to p_{\rm sp}\ pn\ e^+e^-$  at
low values of the spectator momentum, $\bsp=0.25\: \bpD$ (left panel)
and $\bsp=0.35\: \bpD$ (right panel).
Dashed lines depict the  background contribution, dotted lines
are separate contributions from  $\rho$ and $\omega$ mesons, and
solid lines are for the total cross section.}
\label{fig15}
\end{figure}

\begin{figure}[h]  %      Fig17  FSI + VMD in Dp
\includegraphics[width=0.999\textwidth]{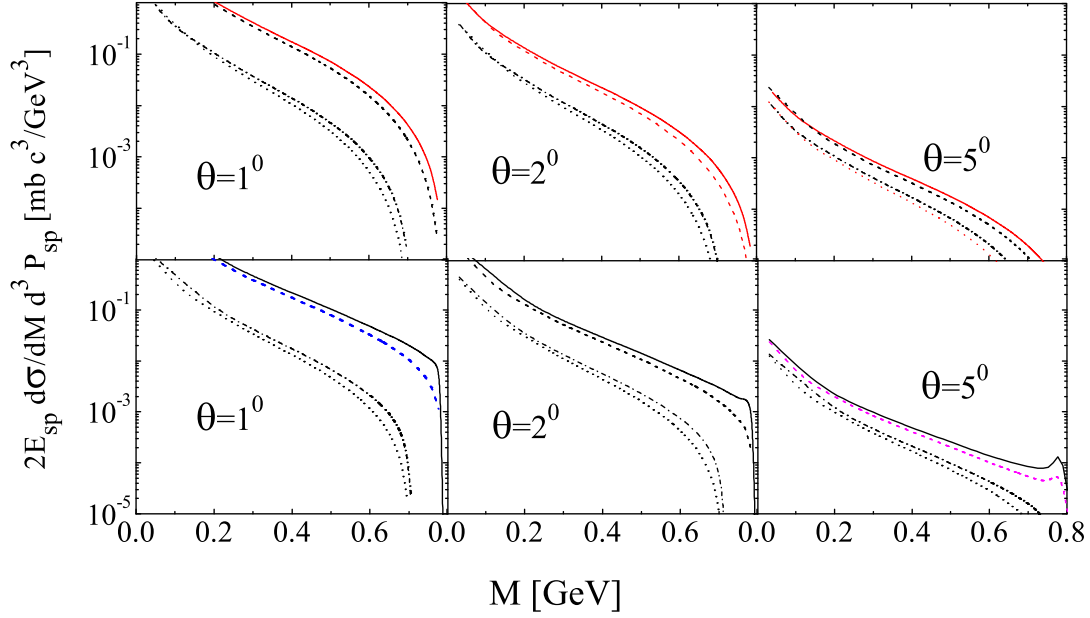} %
\caption{Illustration of FSI effects for
the reaction $Dp\to p_{\rm sp}\ pn \ e^+e^-$  at
$T_{kin}=\:  1.9\: A\  GeV$ and two  values of the spectator momentum.
Solid (dashed) lines correspond to results with (without) FSI taken into account
for the spectator momentum $\bsp=0.5\: \bpD$,
while the  dot-dashed  (dotted) lines correspond
to $\bsp=0.55\: \bpD$, respectively. Calculations have been performed at three
different values of the spectator angle, $\theta=1^\circ$, $2^\circ$ and
$5^\circ$. In the upper row solely contribution from background
is displayed, while in the lower row VMD effects have been included as well.}
\label{fig17}
\end{figure}

\end{document}